\documentclass[12pt]{JHEP3} 

\usepackage{amsmath,amssymb}

\newcommand{\Tr}{{\rm Tr \,}}

\renewcommand{\v}{{\varphi}}
\renewcommand{\o}{{\omega}}

\newcommand{\be}{\begin{equation}}
\newcommand{\ee}{\end{equation}}
\newcommand{\bea}{\begin{eqnarray}}
\newcommand{\eea}{\end{eqnarray}}

\newcommand{\pint}{\makebox[0pt][l]{\hspace{3.4pt}$-$}\int}

\newcommand{\cO}{{\cal O}}
\newcommand{\cN}{{\cal N}}
\newcommand{\cP}{{\cal P}}
\newcommand{\cH}{{\cal H}}
\newcommand{\cC}{{\cal C}}

\newcommand{\atopfrac}[2]{\genfrac{}{}{0pt}{}{#1}{#2}}
\newcommand{\sfrac}[2]{{\textstyle\frac{#1}{#2}}}
\newcommand{\half}{\sfrac{1}{2}}

\newcommand{\contourgauge}{\mathbf{C}}

\newcommand{\alg}[1]{\mathfrak{#1}}
\newcommand{\alSU}{\alg{su}}
\newcommand{\su}{\alg{su}}

\newcommand{\psu}{\alg{psu}}
\newcommand{\sll}{\alg{sl}}
\newcommand{\indups}[1]{_{\mathrm{\scriptscriptstyle #1}}}
\newcommand{\gym}{g\indups{YM}}
\newcommand{\upper}[1]{^{\mathrm{\scriptscriptstyle #1}}}

\title{
The Factorized S-Matrix of CFT/AdS
}

\author{Matthias Staudacher \\
\vskip 0.5cm
Max-Planck-Institut f\"ur Gravitationsphysik,
Albert-Einstein-Institut\footnote{
Permanent Address}\\
Am M\"uhlenberg 1, D-14476 Potsdam, Germany
\vskip 0.1 cm
and
\vskip 0.1 cm
Kavli Institute for Theoretical Physics,
University of California\\
Santa Barbara, CA 93106 USA
\vskip 0.2cm
Email: \email{matthias@aei.mpg.de}   }

\preprint{\hepth{0412188}\\
AEI-2004-107\\
NSF-KITP-04-122}

\abstract{
We argue that the recently discovered integrability in the 
large-$N$ CFT/AdS system is equivalent to diffractionless 
scattering of the corresponding hidden elementary excitations. 
This suggests that, perhaps, the key tool for finding the spectrum of this 
system is neither the gauge theory's dilatation operator nor the 
string sigma model's quantum Hamiltonian, but instead the respective 
factorized S-matrix. To illustrate the idea, we focus on the 
closed fermionic $\su (1|1)$ sector of the $\cN=4$ gauge theory. 
We introduce a new technique, the perturbative asymptotic Bethe 
ansatz, and use it to extract this sector's three-loop S-matrix 
from Beisert's involved algebraic work on the three-loop $\su(2|3)$ 
sector. We then show that the current knowledge about semiclassical
and near-plane-wave quantum strings in the $\su(2)$, $\su(1|1)$ and $\sll(2)$
sectors of $AdS_5 \times S^5$ is fully consistent with the existence 
of a factorized S-matrix. Analyzing the available information,
we find an intriguing relation between the three associated
S-matrices. Assuming that the relation also holds in gauge
theory, we derive the three-loop S-matrix of the $\sll(2)$ sector
even though this sector's dilatation operator is not yet known
beyond one loop. The resulting Bethe ansatz reproduces the 
three-loop anomalous dimensions of twist-two operators recently 
conjectured by Kotikov, Lipatov, Onishchenko and Velizhanin,
whose work is based on a highly complex QCD computation of
Moch, Vermaseren and Vogt.

}


\begin{document}

\section{Introduction and Conclusions}

There is mounting evidence that four-dimensional strictly
planar $\cN=4$ Yang-Mills gauge theory is integrable. 
Likewise, there are strong indications that free IIB 
superstring theory on the curved space $AdS_5 \times S^5$ 
is also integrable. This means that there is hope that
the spectrum of both theories might be exactly computable.
If true, the AdS/CFT duality conjecture becomes 
{\it falsifiable}: Either the spectrum of the two models
agrees, or it does not. This is good news, since, according
to scientific tradition, falsifiability is an important
feature of a {\it theory}, to be distinguished from 
creative speculation.

The first, crucial hint that conformal $\cN=4$ gauge theory 
might be integrable in the planar limit was discovered in a
beautiful paper by Minahan and Zarembo \cite{MZ}.
There it was shown that the spectrum of $\cN=4$ conformal 
operators may be obtained by diagonalizing an integrable quantum 
spin chain. The observation was initially restricted to the subset
of scalar operators at one loop. However, shortly after, strong evidence
was found that integrability extends to higher loops 
\cite{BKS}, and, at least at one loop, to the full set
of $\cN=4$ operators \cite{BS}. Integrable structures
have appeared before in planar QCD, starting
with the pioneering work of Lipatov \cite{lipatov}
(for a comprehensive review, see \cite{review}).
There it was always considered to be due to approximate,
hidden symmetries, which can lead to useful information
on high energy scattering and QCD anomalous dimensions.
In $\cN=4$ \cite{MZ,BKS,BS} the goal is more
ambitious. One would like to exactly {\it solve} a
four-dimensional gauge theory. 

In a parallel, intertwined development integrable structures
were also observed in the worldsheet theory of strings
on $AdS_5 \times S^5$. First, there were some hints that
the relevant classical coset sigma model is integrable
\cite{BPR}. Later it was shown in detail how finite-dimensional
reductions of the classical sigma model lead to
integrable classical mechanics systems of Neumann type 
\cite{AFRT}. The reduction may be undone by the B\"acklund
transformation, and the infinite tower of commuting charges
of the original sigma model is recovered \cite{AS}.
Subsequently, in an
important paper by Kazakov, Marshakov, Minahan and Zarembo,
it was demonstrated how the spectrum of finite-gap solutions of
the classical sigma model may be
obtained by the classical inverse scattering method \cite{KMMZ}.
In \cite{AFS} first steps toward finding the quantum spectrum of the
sigma model were taken, and an approximate S-matrix was proposed.
Very recently, the Hamiltonian of the classical bosonic string
propagating on $AdS_5 \times S^5$ was shown to be integrable by
constructing, in a special gauge, the corresponding Lax representation
\cite{AF}.

How should one proceed toward, for one, exactly solving the planar ``CFT'',
namely the large $N$ limit of $\cN=4$ gauge theory, and, secondly,  
the worldsheet theory of strings on $AdS_5 \times S^5$? 
Let us assume, as a working hypothesis, that both models are
indeed not only approximately, but in fact completely integrable.
If true, proceeding with the construction of the gauge theory's
dilatation operator, loop-by-loop, or meticulously quantizing
the full sigma model, including fermions, might prove too hard.
In fact, integrable systems do not always have simple Hamiltonians.
To the contrary, the Hamiltonian might be quite intricate
in order to realize the subtle hidden symmetries responsible for
integrability. 

Is there a simpler object which encodes the spectrum of
an integrable quantum system? In fact there is: The S-Matrix.
Quantum integrability is deeply tied to the concept of
diffractionless, factorized scattering\footnote{We are
discussing in this paper an {\it internal}   
S-Matrix describing the scattering of elementary excitations
on a lattice hidden inside the trace of of gauge invariant
composite local operators. It should not be confused
with the {\it external} S-matrix of $\cN=4$ which refers to multi-gluon
amplitudes in four-dimensional space time. Recently
dramatic progress was also achieved in this direction,
see \cite{BDK}, and references therein and thereafter.
It would be exciting if a relation between the internal and the
external S-matrix could be found.}. 
It means that the 
elementary excitations of a quantum many-body system interact 
only through a sequence of two-body scattering processes
which may lead to the exchange of quantum numbers and momenta,
but do not alter the magnitudes of the latter.
This is the next-best thing to a free system! Interaction
does take place, but scattering only leads to a 
permutation of a fixed set of momenta (and, possibly, of 
quantum numbers). 

Here we would like to propose that one should try to directly construct the
S-matrix of integrable gauge and string theory. Hopefully, the
two will then agree. Unfortunately we do not have a very concrete
proposal about how to go about it in a direct fashion. 
However, since the AdS/CFT system possesses a huge amount of symmetry,
and, in particular, apparently an infinite number of hidden
conserved charges, we may certainly 
hope that methods to ``bootstrap'' the S-matrix
will eventually be found.

As a more modest first step, we will analyze part of what is
known about the gauge theory's dilatation operator, and about
the spectrum of AdS strings, in order to get a glimpse at
how the S-matrix might look like. We think that the results
are encouraging. For simplicity we will restrict ourselves, 
in this paper, to three simple two-component sectors of
the full superconformal, $\psu(2,2|4)$ symmetric, system. 
These are a compact $\su(2)$ subsector (two bosons),
a compact $\su(1|1)$ subsector (one boson, one fermion), and a
non-compact $\sll(2)$ subsector (two bosons).

Let us recall what is currently known about the $\cN=4$
gauge theory's dilatation operator and its Bethe ansatz.
The latter may be constructed once the S-matrix is known.
At one loop, the dilatation operator was derived for the 
complete set of $\psu(2,2|4)$ operators \cite{complete}.
It was shown to be integrable in \cite{BS}, and the associated
Bethe ansatz was constructed. 
Beyond one loop, we have Beisert's two- and
three loop dilatation operator in the ``maximally compact''
subsector $\su(2|3)$ \cite{dynamic}. This includes the $\su(2)$ 
(here it was initially found in \cite{BKS}) and $\su(1|1)$ sectors, 
but not the $\sll(2)$ sector. No two- and three loop Bethe 
ansatz is known, except for the $\su(2)$ sector, where it was 
constructed in \cite{SS}. The $\su(2)$ dilatation operator was
then extended to five loops, assuming integrability as well
as BMN scaling.
A five-loop Bethe ansatz was found experimentally, and
an all-loop (asymptotic) Bethe ansatz was proposed \cite{BDS}.
For a closely related system, see \cite{KP}.
Note that the three-loop predictions of the three-loop dilatation 
operator in the $\su(2)$ sector were recently spectacularly confirmed,
for two states, by a full-fledged, rigorous field theory
computation \cite{EJS}.
For a detailed review on most aspects concerning the 
$\cN=4$ dilatation operator, see \cite{thesis}.

One thing to note about the higher
loop dilatation operators \cite{BKS,dynamic,BDS}
is that they are {\it very} complicated,
and appear to increase exponentially in complexity as the
loop order increases (see e.g.~appendix A.1 in \cite{BDS}). 
In fact, the $\su(2|3)$ dilatation 
operator \cite{dynamic} is only known in ``algorithmic'' form.  

In contradistinction, the S-matrices appear to allow for a much
simpler description. E.g.~the (asymptotic) S-matrix of 
\cite{BDS} may be written in a very compact fashion 
({\it cf} \eqref{bds} with \eqref{pf}). We shall see
another example in this paper, where we extract the three-loop
S-matrix of the $\su(1|1)$ subsector from \cite{dynamic}.
This requires designing a new technique which one might term 
PABA: Perturbative Asymptotic Bethe Ansatz. To motivate
it, we derive in some, hopefully pedagogical, detail various
one-loop Bethe ans\"atze at the beginning of the paper.
The final result for $\su(1|1)$ is rather compact, while the ``source'',
namely the two- and three-loop dilatation operator, fills
at least a page. Clearly things will get worse for the operator
at even higher loops, while we suspect the existence of a simple 
all-loop expression for the S-matrix.

In light of the AdS/CFT correspondence it is very natural
to ask about the S-matrix of the string sigma model.
However, the sigma model believed to describe strings on
$AdS_5 \times S^5$ is far more complicated than those theories
where S-matrices have already been found. Still, one could try to
look for similar models in order to derive some hints,
which is an approach that has recently been followed in 
\cite{MP}.

Alternatively, one can attempt to derive useful clues
from available information on the sigma model's spectrum. 
It comes from two sources. We already mentioned 
\cite{KMMZ}, where an equation describing the classical
spectrum in the $\su(2)$ sector (in gauge theory connotation)
was derived. Recently, the procedure was also applied to
the $\sll(2)$ sector \cite{KZ}. It is found that the most
general classical finite-gap solution may be described by
an algebraic curve. But does it contain
information on the S-matrix, which is an inherently
quantum concept? Here it turns out that the crucial
connection does not come from the curve as such, but
from the {\it equation} describing it. Namely, the 
equation may be brought into a ``scattering form''
which allows to draw conclusions about the interactions
of the {\it quantum} excitations as one approaches the
classical limit \cite{BDS},\cite{AFS}. The crucial 
intuition for finding the correct interpretation comes from the fact
that at one- and two-loops the classical string sigma model
behaves very similarly \cite{FT,AFRT} (but, starting from
three loops, not identical \cite{SS}) to the gauge
theory when the associated operators become very ``long''.
On the gauge side one finds rather similar equations
in this ``thermodynamic'' situation \cite{BMSZ,BFST},
\cite{SS}. Since the latter, which are derived from the
discrete Bethe equations, do have a scattering interpretation,
the detailed comparison allows one to also bring the
string equations into ``scattering form''. This logic gave a hint
about the string's S-matrix in the $\su(2)$ sector \cite{AFS},
and we will apply it below, using \cite{KZ}, to
learn about the S-matrix in the $\sll(2)$ sector.

A second important source of spectral information comes
from the plane wave limit of strings on $AdS_5 \times S^5$,
where the worldsheet theory becomes free, and may be thus be quantized 
exactly \cite{METS}. Much intuition about the nature
of the elementary excitations in this limit comes from
subsequently comparing to gauge theory, i.e.~from considering
the famous BMN limit \cite{BMN}. 
Recent studies of the near-BMN limit 
\cite{PR,Call1,Call2,CHMS,McLS} 
yield information on how these excitations begin to interact.
As we will show below, these interactions are fully
consistent with the idea of an underlying factorized S-matrix.
In fact, we will use the results of 
\cite{Call2,CHMS,McLS} to derive part of the
string S-matrix in the {\it fermionic} sector 
$\su(1|1)$. This is important, as it is unclear what
may be learned from the classical sigma model for this
sector. (For some recent work in this direction, see \cite{esperanza}.) 
After all, fermions are inherently quantum.
They will actually serve as a  ``missing link'' in the present 
story \ldots

\ldots which finishes as follows. Comparing the approximate
S-matrices extracted from string theory in the three sectors
$\su(2)$, $\su(1|1)$ and $\sll(2)$, we find a simple 
equation relating them. We then apply the relation to
three-loop gauge theory, and write down the S-matrix in
the $\sll(2)$ sector, even though we do not know the
higher loop dilatation operator. The resulting spectrum
is then successfully checked in the case of two-impurity states,
where the spectrum is known from superconformal symmetry
\cite{2imp}. More excitingly, we may compare to recent
results of Kotikov, Lipatov, Onishchenko and Velizhanin 
\cite{KLOV}, who proposed three-loop exact anomalous
dimensions for $\cN=4$ twist-two operators. 
These dimensions were first found at one-loop from the OPE of four-point
functions in \cite{DO}, and the first few, at two loops,
in \cite{AEPS}. This author does not understand 
the derivation of Kotikov et.al., but it is clear
that the crucial input is a computationally highly
intensive three-loop QCD field theory computation
recently completed after a many-year effort 
by Moch, Vermaseren and Vogt \cite{MVV}
(see also the comments made in \cite{BDK}).
It involved the evaluation of $\cO(10^6)$ auxiliary integrals, 
and the development of cutting-edge algorithmic techniques. 
The spectrum of twist-two operators
derived from our S-matrix {\it agrees} with the results
of Kotikov et.al.

Can we do better and find the full S-matrix of the CFT and
strings on AdS in a more direct fashion? Note that
this possibility entirely hinges on whether integrability
is really exact in either theory. It does {\it not} necessarily
depend on whether the AdS/CFT duality \cite{M} is actually valid dynamically. 
We feel that the present work demonstrates once again
that it is surely useful.


\section{One-Loop Scattering in Planar $\cN=4$}
\label{oneloop}

\subsection{The $\su(2)$ Bosonic Sector}
\label{su2}

This sector consists of operators of the type
\bea\label{su2ops}
\Tr \phi^M Z^{L-M}+ \ldots = \Tr \phi^M Z^J+ \ldots\, ,
\eea
where $J$ denotes an R-charge w.r.t.~SO(6) and $M$ the number of
``impurities''. The partons $Z$ and $\phi$ are two out of the
three complex adjoint scalars of the $\cN=4$ model. 
The dots indicate that we need to consider all possible
orderings of them inside the trace, and
diagonalize the set of such operators with respect to dilatation.
This is most easily done when interpreting the dilatation
operator as a spin chain Hamiltonian \cite{MZ}. 
In the spin chain interpretation 
$L$ is the chain length and $M$ the number of excitations. 
It is convenient to open up the trace and replace it
by a quantum mechanical state on a one dimensional lattice of $L$ sites:
\bea\label{replace}
\Tr \left( \phi Z Z \phi \ldots \phi Z \right)
\rightarrow
|\phi Z Z \phi \ldots \phi Z \rangle.
\eea
Let us label the sites of this lattice by a discrete coordinate $x$.
(So in the example of \eqref{replace} we see that we have a $\phi$ at 
$x=1,4,L-1$ and a $Z$ at $x=2,3,L$). The fact that we originally 
had a trace leads to {\it two} consequences, to be distinguished.
The first is that, since the trace links the matrix indices of
the first and last parton, the chain has periodic boundary conditions:
$x=L+1$ is to be identified with $x=1$. The second
is that the cyclicity of the trace requires us to 
project onto states whose total lattice momentum is zero, see 
\eqref{mom} below.

The planar one-loop position space Hamiltonian reads
\bea
H_0=\sum_{x=1}^L\, (1-\cP_{x,x+1})=
\sum_{x=1}^L\,
\half (1-\vec\sigma_x \cdot \vec\sigma_{x+1})\, ,
\eea
and may be expressed with the help of the permutation operator 
$\cP_{x,x+1}$ which exchanges the partons at sites $x$ and $x+1$,
as first noticed in \cite{MZ,BKPS}. 
It may alternatively be interpreted as
an $\su(2)$ nearest neighbor spin chain \cite{MZ} 
if we say that $Z$ is a spin up with 2-spinor $(1,0)$ and
$\phi$ is a spin down with 2-spinor $(0,1)$.  
Then the spin operator at lattice 
site $x$ contains the three Pauli matrices 
$\vec \sigma_x=(\sigma_x^1,\sigma_x^2,\sigma_x^3)$ and the
Hamiltonian may be written in the the form of a sum over
nearest neighbor spin-spin interactions.

The two-body states are defined by
\bea
\label{su2state}
\atopfrac{{ }~~~~~~~~~~~~~~~~~~~~~~~~~~~~~~~~~~~~~\atopfrac{x_1}{\downarrow}
~~~~~~\atopfrac{x_2}{\downarrow}}
{|\Psi\rangle=\sum_{1\leq x_1 < x_2\leq L} \Psi(x_1,x_2)~
| ... Z \phi Z ...Z \phi Z...\rangle\, ,}
\eea
where $x_{1,2}$ (with $x_1<x_2$) label the positions of the two $\phi$ particles
in the background of the $Z$ particles. In position space the
Schr\"odinger equation $H_0 \cdot |\Psi\rangle=E_0\,|\Psi\rangle$
becomes
\bea\label{su2schroedingerO}
{\rm for} \quad x_2> x_1+1 : & & \\ 
E_0\,\Psi(x_1,x_2)&=& 
2\,\Psi(x_1,x_2)-\Psi(x_1-1,x_2)-\Psi(x_1+1,x_2)+
\nonumber \\
& &+2\,\Psi(x_1,x_2)-\Psi(x_1,x_2-1)-\Psi(x_1,x_2+1)\, ,
\nonumber  \\ \label{su2schroedingerII}
{\rm for} \quad  x_2= x_1+1:
 \\
E_0\,\Psi(x_1,x_2)&=& 
2\,\Psi(x_1,x_2)-\Psi(x_1-1,x_2)-\Psi(x_1,x_2+1)\, .
\nonumber 
\eea

This difference equation is easily solved by Bethe's ansatz
\cite{Bethe} 
for the position space wave function $\Psi(x_1,x_2)$ which reads
\bea\label{su2ansatz}
\label{oneloopbethe}
\Psi(x_1,x_2)=e^{i p_1 x_1+i p_2 x_2}+
S(p_2,p_1)~e^{i p_2 x_1+i p_1 x_2}\, .
\eea
It is based on the intuition that the partons should freely evolve
down the trace with fixed momenta $p_1$,$p_2$ until they hit
each other at $x_2=x_1+1$. If the system is integrable they
should then simply either pass through each other, or else exchange
momenta, with an amplitude given by the S-matrix $S(p_1,p_2)$.
This scattering process is {\it non-diffractive} if the individual
momenta $p_k$ are individually conserved. If we were to stay with
the two-body problem this would of course always be true because
of total momentum conservation. The miracle of integrability
is that it remains true for the $M$-body problem. Put differently,
we cannot prove integrability by only considering the two-body problem,
but we can certainly find the S-matrix if we assume (or know)
that the system is integrable.

Plugging the ansatz \eqref{su2ansatz} into the Schr\"odinger
equation \eqref{su2schroedingerO},\eqref{su2schroedingerII} 
one finds that it is indeed
satisfied if, firstly, the energy is given by the 
dispersion law
\bea\label{1Ldispersion}
E_0=\sum_{k=1}^M\, 4\, \sin^2\left(\frac{p_k}{2}\right)\, ,
\eea
with $M=2$, and, secondly, if the S-matrix is given by
\bea\label{su2smatrix}
S_{\su(2)}(p_1,p_2)=
-\frac{e^{i p_1+i p_2}-2 e^{i p_1}+1}{e^{i p_1+i p_2}-2 e^{i p_2}+1}\, .
\eea
In line with intuition, the dispersion law \eqref{1Ldispersion}
follows from the ``generic'' situation \eqref{su2schroedingerO}
while the S-matrix is deduced from the ``colliding'' 
situation \eqref{su2schroedingerII}.

All this is true for arbitrary values of the momenta $p_k$. 
As always in quantum mechanics the eigenvalues get fixed
through the boundary conditions. Imposition of periodic boundary 
conditions $\Psi(x_1,x_2)=\Psi(x_2,x_1+L)$
on the wave function \eqref{su2ansatz} leads to
Bethe's equations:
\bea
\label{twoimpbethe}
e^{i p_1 L}=S(p_1,p_2) \qquad {\rm and} \qquad e^{i p_2 L}=S(p_2,p_1)\, .
\eea
Their solution leads in many cases to complex values of the
momenta $p_k$. This simply reflects the fact that
the partons are not freely propagating, but mutually interacting 
inside a finite volume. In particular, they can form bound states.

The principle of non-diffractive scattering, if applicable,
now allows us to take a big leap and immediately solve
the $M$-body problem. The total phase factor acquired by a parton 
circling around the trace should simply be given as a product of the
phase factors due to individual collisions with all other
$M-1$ partons. The Bethe equations become
\bea
\label{bethe}
e^{i p_k L}=
\prod_{\textstyle\atopfrac{j=1}{j\neq k}}^M\, S(p_k,p_j)\, ,
\qquad \qquad
k=1,\ldots,M\, ,
\eea
with the {\it same} two-body S-matrix \eqref{su2smatrix}!
The total energy is of course still given by the sum
over all $M$ local dispersion relations \eqref{1Ldispersion}.

In our gauge theory application we have to take
into account the fact that the trace is cyclic, which means that
we are only interested in the zero-momentum sector:
\bea\label{mom}
\sum_{k=1}^M\,p_k=0\, .
\eea
The one-loop anomalous dimension $\Delta$ of the original gauge theory
operators \eqref{su2ops} is then related to the energy spectrum of the 
zero-momentum states of the nearest-neighbor spin chain
through
\bea\label{Delta}
\Delta=\Delta_0+g^2\, E_0+\cO(g^4)\, ,
\qquad {\rm with} \qquad 
g^2=\frac{\gym^2 N}{8 \pi^2}\, ,
\eea
where in the present case the classical scaling dimension 
is $\Delta_0=L$.

Finally, one commonly expresses the $\su(2)$ Bethe equations in a form
familiar from the algebraic Bethe ansatz \cite{faddeev}, 
as in \cite{MZ,BS}.
Changing variables by introducing the so-called Bethe roots 
$u_k=\half \cot\left(\frac{p_k}{2}\right)$,
\eqref{bethe} becomes, after rewriting the S-matrix \eqref{su2smatrix},
\bea
\left(\frac{u_k+\frac{i}{2}}{u_k-\frac{i}{2}}\right)^L=
\prod_{\textstyle\atopfrac{j=1}{j\neq k}}^M\,
\frac{u_k-u_j+i}{u_k-u_j-i}\, ,
\qquad \qquad
k=1,\ldots,M\, ,
\eea
while the momentum constraint \eqref{mom} and the energy 
\eqref{1Ldispersion} turn into
\bea\label{umom}
\prod_{k=1}^M\,\frac{u_k+\frac{i}{2}}{u_k-\frac{i}{2}}=1
\qquad {\rm and} \qquad
E_0=\sum_{k=1}^M\, \frac{1}{u_k^2+\frac{1}{4}}\, .
\eea

\subsection{The $\su(1|1)$ Fermionic Sector}
\label{psu11}

This sector consists of operators of the type
\bea\label{psu11ops}
\Tr \psi^M Z^{L-M}+ \ldots = \Tr \psi^M Z^{J-\frac{M}{2}}+ \ldots\, ,
\eea
where $J$ denotes an R-charge w.r.t.~SO(6) and $M$ the number of
``impurities''. The partons are for one a complex adjoint scalars $Z$,
and secondly an adjoint gaugino $\psi$ (in $\cN=1$ connotation).  
In the spin chain interpretation 
$L$ is the chain length and $M$ the number of excitations.

The planar one-loop Hamiltonian reads
\bea
H_0=\sum_{x=1}^L \,(1-\Pi_{x,x+1})\, .
\eea
It may be extracted from the complete one-loop $\cN=4$ 
dilatation operator \cite{complete} and rewritten
with the help of the graded permutation operator $\Pi_{x,x+1}$
which exchanges the partons at sites $x$ and $x+1$, picking
up a minus sign if the exchange involves two fermions $\psi$.
It was noticed in \cite{CHMS} that this Hamiltonian corresponds
to a {\it free} lattice fermion. Here we would like to put this
observation into a familiar condensed matter context.
Let us rewrite the Hamiltonian in spin chain form by expressing it
with the help of the three Pauli matrices 
$\sigma_x^1,\sigma_x^2,\sigma_x^3$ :
\bea\label{xy}
H_0=
\sum_{x=1}^L\,\Big(
(1-\sigma_x^3)
-\half (\sigma_x^1 \sigma_{x+1}^1+\sigma_x^2 \sigma_{x+1}^2)
\Big)\, .
\eea
In spin chain language
the bosonic partons $Z$ are ``spin up'' spinors $(1,0)$ and the
fermionic partons $\psi$ are ``spin down'' spinors $(0,1)$.
We now observe that \eqref{xy} is the Hamiltonian of    
an XY spin chain in a magnetic field, which is well known to 
correspond to free lattice fermions. It is therefore
even simpler than the XXX Heisenberg model! 
(The model is isotropic in the $\sigma_x^1,\sigma_x^2$ plane,
so we could call it an XX spin chain.) 

The two-body states are defined by
\bea
\label{psu11state}
\atopfrac{{ }~~~~~~~~~~~~~~~~~~~~~~~~~~~~~~~~~~~~~\atopfrac{x_1}{\downarrow}
~~~~~~\atopfrac{x_2}{\downarrow}}
{|\Psi\rangle=\sum_{1\leq x_1 < x_2\leq L} \Psi(x_1,x_2)~
| ... Z \psi Z ...Z \psi Z...\rangle\, ,}
\eea
where $x_1$,$x_2$ label the positions of the two gauginos in the
background of the bosonic $Z$ particles. In position space the
Schr\"odinger equation $H_0 \cdot |\Psi\rangle=E_0\,|\Psi\rangle$
becomes
\bea\label{psu11schroedingerO}
{\rm for} \quad x_2> x_1+1 : & & \\ 
E_0\,\Psi(x_1,x_2)&=& 
2\,\Psi(x_1,x_2)-\Psi(x_1-1,x_2)-\Psi(x_1+1,x_2)+
\nonumber \\
& &+2\,\Psi(x_1,x_2)-\Psi(x_1,x_2-1)-\Psi(x_1,x_2+1)\, ,
\nonumber  \\ \label{psu11schroedingerII}
{\rm for} \quad  x_2= x_1+1:
 \\
E_0\,\Psi(x_1,x_2)&=& 
4\,\Psi(x_1,x_2)-\Psi(x_1-1,x_2)-\Psi(x_1,x_2+1)\, .
\nonumber 
\eea
These equations are identical to the ones of the
$\su(2)$ model except for the innocent looking replacement
of a factor of 2 by a 4 when comparing 
\eqref{su2schroedingerII} and \eqref{psu11schroedingerII}.

We now make the same Bethe ansatz \eqref{oneloopbethe}
as in the case of the $\su(2)$ sector, and plug it into
the difference Schr\"odinger equations 
\eqref{psu11schroedingerO},\eqref{psu11schroedingerII}.
The energy eigenvalue \eqref{1Ldispersion} remains unchanged,
but the result for the S-matrix is very different:
\bea\label{psu11smatrix}
S_{\su(1|1)}(p_1,p_2)=-1\, .
\eea
So indeed the excitations behave as free fermions, and
the S-matrix \eqref{psu11smatrix} reflects free fermi statistics.
In fact, here the wavefunction \eqref{oneloopbethe} is simply a 
two-body Slater determinant. Some care has to be taken with
Fermi statistics when imposing periodic boundary 
conditions; we now have $\Psi(x_1,x_2)=-\Psi(x_2,x_1+L)$.
The ``Bethe equations'' are then:
\bea
\label{twoimpfermibethe}
e^{i p_1 L}=-S(p_1,p_2)=1 \qquad {\rm and} \qquad e^{i p_2 L}=-S(p_2,p_1)=1\, .
\eea
Due to integrability we can again immediately solve
the one-loop $M$-body problem. Here there simply is no
scattering at all, but we need to be careful with Fermi statistics.
The $M$-body equations read
\bea
\label{psu11bethe}
e^{i p_k L}=1\, ,
\qquad \qquad
k=1,\ldots,M\, .
\eea
These are of course immediately solved
\bea\label{freefermion}
p_k=\frac{2 \pi n_k}{L}\, ,
\eea
but due to fermi statistics the integer mode number $n_k$
are required to be all distinct. (The wavefunction is
an $M\times M$ Slater determinant and vanishes if two momenta
coincide).
They also have to be restricted to the fundamental
Brillouin zone in order to avoid overcounting the states.
Furthermore, in gauge theory we are again only interested in the 
zero-momentum sector $\sum_{k=1}^M\,p_k=0$, which immediately
translates into $\sum_{k=1}^M\,n_k=0$.

The one-loop anomalous dimension $\Delta$ of the original gauge theory
operators \eqref{psu11ops} is still given by \eqref{Delta} where
the classical scaling dimension is $\Delta_0=L+\half\, M$.

\subsection{The $\sll(2)$ Derivative Sector}
\label{sl2}

This sector consists of operators of the type
\bea\label{sl2ops}
\Tr D^M Z^L+ \ldots = \Tr D^S Z^J+ \ldots\, ,
\eea
where $J$ denotes an R-charge w.r.t.~SO(6) and $M=S$ the number of
``impurities'' which corresponds to one of the two spin quantum 
numbers\footnote{We will mostly avoid using the customary letter $S$ for
the AdS spin in this paper in order to prevent, however unlikely, 
confusions with the S-matrix. Instead, we will use $M$ 
(for {\it magnon} number).} $S$
of SO(2,4). The partons are for one a complex adjoint scalar $Z$,
and secondly an adjoint lightcone covariant derivative $D$.
The dots indicate that we need to consider all possible
distributions of the covariant derivatives $D$ onto the scalars  
$Z$ inside the trace, and diagonalize the set of such operators with 
respect to dilatation. 
In the spin chain interpretation 
$L=J$ is the chain length and $M$ the number of excitations.
Note that, unlike the previous two cases, the excitations do not
contribute to the length of the spin chain. 
In particular, the number of excitations may exceed the total
length of the chain. In this context, the length is also 
commonly called ``twist''. A ``twist-two'' operator is 
thus a very short spin chain of length 2.

The one-loop Hamiltonian for this sector has a global
$\sll(2)$ invariance. It is non-polynomial, and has been derived 
and described in \cite{complete}.
In the planar limit it may be considered as an integrable nearest neighbor
$\sll(2)$ spin $-\half$ spin chain \cite{BS}. Let us express the Hamiltonian
$H$ through the Hamiltonian density $\cH$: 
\bea
H_0= \sum_{x=1}^L\,\cH_{x,x+1}\, .
\eea
Denoting partons with $0,1,2$ derivatives, 
i.e.~$Z$, $DZ$, $D^2Z$ by, respectively,  
$|0\rangle$, $|1\rangle$, $|2\rangle$, we can read off the action of
the Hamiltonian density on a neighboring pair of partons 
at lattice sites $x,x+1$ from \cite{complete} as
\bea
\begin{array}{c}
\displaystyle
\cH \cdot |1,0\rangle =|1,0\rangle-|0,1\rangle 
\hspace*{1cm}
\cH \cdot |0,1\rangle =|0,1\rangle-|1,0\rangle 
\\ [0.3cm]
\displaystyle
\cH  \cdot |1,1\rangle =2\,|1,1\rangle-|2,0\rangle-|0,2\rangle
\\ [0.3cm]
\displaystyle
\cH \cdot |2,0\rangle =
\frac{3}{2}\, |2,0\rangle-|1,1\rangle -\frac{1}{2}\, |0,2\rangle
\hspace*{1cm}
\cH \cdot |0,2\rangle =
\frac{3}{2}\, |0,2\rangle-|1,1\rangle -\frac{1}{2}\, |2,0\rangle\, ,
\end{array}
\eea
where we have abbreviated $\cH_{x,x+1}$ by $\cH$.
We have not written any terms beyond the two-body interactions as 
we shall not need them for finding the S-matrix. They {\it are} of
course important for proving the factorizability of the many-body
S-matrix, but {\it not} for finding it once we know (or believe)
that the model is integrable. This observation
might be very useful when constructing the Bethe ansatz
for this and other sectors at higher loops, as it should apply 
there as well. It also illustrates nicely one of the main points
of this paper, namely that the S-matrix is a much simpler object
than the dilatation operator!

The two-body states are defined by
\bea
\label{Dstate}
\atopfrac{{ }~~~~~~~~~~~~~~~~~~~~~~~~~~~~~~~~~~~~~\atopfrac{x_1}{\downarrow}
~~~~~~~~~~~\atopfrac{x_2}{\downarrow}}
{|\Psi\rangle=\sum_{1\leq x_1 \leq x_2\leq L} \Psi(x_1,x_2)~
| ... Z (D Z) Z ...Z (D Z)Z ...\rangle\, ,} 
\eea
where $x_1$, $x_2$ label the positions of the two derivatives in
the background of the bosonic $Z$ particles.
Notice the following subtle difference as compared to 
\eqref{su2state},\eqref{psu11state}. The ``particles''
$D$ are not occupying their own lattice sites, but sit instead
on top of the inert particles (=holes) $Z$. In particular, there may
be multiple-occupancy, thus now $x_1=x_2$ is an allowed configuration!
In condensed matter language we would say that the scattering
is not ``hard-core''.  

In position space the Schr\"odinger equation 
$H_0 \cdot |\Psi\rangle=E_0\,|\Psi\rangle$ becomes 
\bea\label{sl2schroedingerO}
{\rm for} \quad x_2> x_1 : & & \\ 
E_0\,\Psi(x_1,x_2)&=& 
2\,\Psi(x_1,x_2)-\Psi(x_1-1,x_2)-\Psi(x_1+1,x_2)+
\nonumber \\
& &+2\,\Psi(x_1,x_2)-\Psi(x_1,x_2-1)-\Psi(x_1,x_2+1)\, ,
\nonumber  \\ \label{sl2schroedingerI}
{\rm for} \quad  x_2= x_1:
 \\
E_0\,\Psi(x_1,x_2)&=&
\frac{3}{2}\, \Psi(x_1,x_2)-\Psi(x_1-1,x_2)-\frac{1}{2}\, \Psi(x_1-1,x_2-1)+
\nonumber \\
& &+\frac{3}{2}\, \Psi(x_1,x_2)-\Psi(x_1,x_2+1)
-\frac{1}{2}\, \Psi(x_1+1,x_2+1)\, .
\nonumber
\eea
Notice that the first of these equations, \eqref{sl2schroedingerO},
is valid for both the ``generic'' situation $x_2>x_1+1$ as well as for
the nearest neighbor situation $x_2=x_1+1$. The second expression
describes the ultralocal on-site interaction $x_2=x_1$ of the
partons.

We make the same Bethe ansatz \eqref{oneloopbethe}
as in the previous two cases. Plugging it into the
first of the two Schr\"odinger difference equations 
\eqref{sl2schroedingerO} gives as before the energy eigenvalue
\eqref{1Ldispersion}, while the equation 
\eqref{sl2schroedingerI}, describing the collision, yields the S-matrix :
\bea\label{sl2smatrix}
S_{\sll(2)}(p_1,p_2)=
-\frac{e^{i p_1+i p_2}-2 e^{i p_2}+1}{e^{i p_1+i p_2}-2 e^{i p_1}+1}\, .
\eea
Notice that it differs from the $\su(2)$ case by the
exchange $p_1 \leftrightarrow p_2$. As before, the imposition
of periodic boundary conditions, together with the principle
of diffractionless scattering, leads to the Bethe equations
of the form \eqref{bethe}. 

Changing once again variables from the momenta $p_k$ to the Bethe roots 
$u_k=\half \cot\left(\frac{p_k}{2}\right)$,
\eqref{bethe} becomes
\bea\label{sl2betheu}
\left(\frac{u_k+\frac{i}{2}}{u_k-\frac{i}{2}}\right)^L=
\prod_{\textstyle\atopfrac{j=1}{j\neq k}}^M\,
\frac{u_k-u_j-i}{u_k-u_j+i}\, ,
\qquad \qquad
k=1,\ldots,M\, ,
\eea
while the momentum constraint and the energy 
are given as in the $\su(2)$ case by \eqref{umom}.

The one-loop anomalous dimension $\Delta$ of the original gauge theory
operators \eqref{sl2ops} is again given by \eqref{Delta} where now
the classical scaling dimension is $\Delta_0=L+M$.

\subsection{Embedding of the Previous Sectors into 
$\su(2,2|4)$}

In the previous sections we demonstrated how to find the 
S-matrix of the three simplest two-component sectors of the
superspin chain of \cite{BS} from the key principle of non-diffractive
scattering. In the next chapter we will show, in the concrete example 
of the fermionic sector $\su(1|1)$, how to extend the method 
to higher loops. This requires the introduction of
a new technique (the perturbative asymptotic Bethe ansatz). 
The method may also be applied to more than two
components, but gets considerably more involved, as one needs to
apply the so-called nested Bethe ansatz \cite{preparation}.
This is beyond the scope of the present paper. 
However, here we would like to show how, at one-loop, the above three 
two-component sectors are recovered as special cases embedded into 
the complete $\su(2,2|4)$ one-loop Bethe ansatz of \cite{BS}.

In the complete super spin chain we have seven types of roots.
It is useful to visualize them with the help of a $\su(2,2|4)$
Dynkin diagram. There are several possible choices, and 
here we will consider the ``Beauty'' version of \cite{BS}:
\bea\label{eq:BeautyRoots}
\begin{minipage}{260pt}
\setlength{\unitlength}{1pt}%
\small\thicklines%
\begin{picture}(260,55)(-10,-30)
\put(  0,00){\circle{15}}%
\put(  0,15){\makebox(0,0)[b]{$ $}}%
\put(  0,-15){\makebox(0,0)[t]{$M_1$}}%
\put(  7,00){\line(1,0){26}}%
\put( 40,00){\circle{15}}%
\put( 40,15){\makebox(0,0)[b]{$ $}}%
\put( 40,-15){\makebox(0,0)[t]{$M_2$}}%
\put( 47,00){\line(1,0){26}}%
\put( 80,00){\circle{15}}%
\put( 80,15){\makebox(0,0)[b]{ }}%
\put( 80,-15){\makebox(0,0)[t]{$M_3$}}%
\put( 87,00){\line(1,0){26}}%
\put(120,00){\circle{15}}%
\put(120,15){\makebox(0,0)[b]{$+1$}}%
\put(120,-15){\makebox(0,0)[t]{$M_4$}}%
\put(127,00){\line(1,0){26}}%
\put(160,00){\circle{15}}%
\put(160,15){\makebox(0,0)[b]{ }}%
\put(160,-15){\makebox(0,0)[t]{$M_5$}}%
\put(167,00){\line(1,0){26}}%
\put(200,00){\circle{15}}%
\put(200,15){\makebox(0,0)[b]{ }}%
\put(200,-15){\makebox(0,0)[t]{$M_6$}}%
\put(207,00){\line(1,0){26}}%
\put(240,00){\circle{15}}%
\put(240,15){\makebox(0,0)[b]{ }}%
\put(240,-15){\makebox(0,0)[t]{$M_7$}}%
\put( 35,-5){\line(1, 1){10}}%
\put( 35, 5){\line(1,-1){10}}%
\put(195,-5){\line(1, 1){10}}%
\put(195, 5){\line(1,-1){10}}%
\end{picture}
\end{minipage}
\eea
On top of the Dynkin diagram we have indicated the
Dynkin labels of the representation corresponding to
the vacuum state of the ``Beauty'' description (which corresponds
to declaring the complex scalars $Z$ to be empty lattice sites or 
``holes''),  and on the bottom the number $M_k$ of roots corresponding to the 
respective node. This way the vacuum state of a length $L$
chain is BPS. The full equations, which can be found in
\cite{BS}, may be concisely written
using the Cartan matrix corresponding to this Dynkin diagram.

The $\su(2)$ $M$-magnon sector is of course immediately obtained by
only exciting the central node: $M_k=\delta_{k4}\, M$.
Now, for the fermionic $\su(1|1)$ chain of section \ref{psu11} 
is not hard to verify \cite{BS} that it corresponds to
the excitation pattern
\bea\label{psu11roots}
\begin{minipage}{260pt}
\setlength{\unitlength}{1pt}%
\small\thicklines%
\begin{picture}(260,55)(-10,-30)
\put( 40,00){\circle{15}}%
\put( 40,15){\makebox(0,0)[b]{$ $}}%
\put( 40,-15){\makebox(0,0)[t]{$M-2$}}%
\put( 40,-30){\makebox(0,0)[t]{$w_l$}}%
\put( 47,00){\line(1,0){26}}%
\put( 80,00){\circle{15}}%
\put( 80,15){\makebox(0,0)[b]{ }}%
\put( 80,-15){\makebox(0,0)[t]{$M-1$}}%
\put( 80,-30){\makebox(0,0)[t]{$v_j$}}%
\put( 87,00){\line(1,0){26}}%
\put(120,00){\circle{15}}%
\put(120,15){\makebox(0,0)[b]{$+1$}}%
\put(120,-15){\makebox(0,0)[t]{$M$}}%
\put(120,-30){\makebox(0,0)[t]{$u_k$}}%
\put( 35,-5){\line(1, 1){10}}%
\put( 35, 5){\line(1,-1){10}}%
\end{picture}
\end{minipage}
\eea
where we have omitted nodes that do not carry excited roots,
and have introduced the notation $u_k$, $v_j$ and $w_l$
for the three types of roots.
The Bethe equations of \cite{BS} become
\bea\label{psu11nested}
\left(\frac{u_k+\frac{i}{2}}{u_k-\frac{i}{2}}\right)^L
&=&
\prod_{\textstyle\atopfrac{k'=1}{k'\neq k}}^M\,
\frac{u_k-u_{k'}+i}{u_k-u_{k'}-i}
\prod_{j=1}^{M-1}\,
\frac{u_k-v_j-\frac{i}{2}}{u_k-v_j+\frac{i}{2}}\, 
\qquad
1\leq k \leq M\, ,
\\
1&=&
\prod_{k=1}^{M}\,
\frac{v_j-u_k-\frac{i}{2}}{v_j-u_k+\frac{i}{2}}
\prod_{\textstyle\atopfrac{j'=1}{j'\neq j}}^{M-1}\,
\frac{v_j-v_{j'}+i}{v_j-v_{j'}-i}
\prod_{l=1}^{M-2}\,
\frac{v_j-w_l-\frac{i}{2}}{v_j-w_l+\frac{i}{2}}\, 
\quad
1\leq j \leq M-1\, ,
\nonumber \\
1&=&
\prod_{j=1}^{M-1}\,
\frac{w_l-v_j-\frac{i}{2}}{w_l-v_j+\frac{i}{2}}\, 
\quad
1\leq l \leq M-2\, .
\nonumber
\eea
Notice the absence of ``self-interactions'' of the 
fermionic roots $w_l$. This allows us to eliminate\footnote{
The following arguments were also independently discovered
by K.~Zarembo (private communication) \cite{Z}.
}
them in the present situation with the following argument.
Introduce 
\bea
q(w):=\prod_{j=1}^{M-1}\,\left(w-v_j+\frac{i}{2}\right)-
\prod_{j=1}^{M-1}\,\left(w-v_j-\frac{i}{2}\right)\, .
\eea
$q(w)$ is clearly a polynomial of degree $M-2$ in $w$.
It therefore has $M-2$ algebraic roots, which, in light of
the last of the above Bethe equations, are precisely the 
fermionic roots $w_l$. We may therefore also write
\bea
q(w)=i\, (M-1)\,\prod_{l=1}^{M-2}\,\left(w-w_l\right)\, .
\eea
This allows us to deduce that for all $1\leq j\leq (M-1)$ 
\bea
\prod_{l=1}^{M-2}\,
\frac{v_j-w_l-\frac{i}{2}}{v_j-w_l+\frac{i}{2}}=
\frac{q(v_j-\frac{i}{2})}{q(v_j+\frac{i}{2})}=
\prod_{\textstyle\atopfrac{j'=1}{j'\neq j}}^{M-1}\,
\frac{v_j-v_{j'}-i}{v_j-v_{j'}+i}\, .
\eea
This however means that the second set of Bethe equations
in \eqref{psu11nested} simplifies significantly:
\bea
1&=&
\prod_{k=1}^{M}\,
\frac{v_j-u_k-\frac{i}{2}}{v_j-u_k+\frac{i}{2}}\, .
\eea
What has happened is that the fermionic roots $w_l$ completely
screen the ``self-interaction'' terms  of the bosonic roots $v_j$, such
that these become fermionic. Clearly this is a general
phenomenon (for this particular excitation pattern) and could
be described pictorially as: 
\bea\label{rootreduction}
\begin{minipage}{260pt}
\setlength{\unitlength}{1pt}%
\small\thicklines%
\begin{picture}(260,55)(-10,-30)
\put( 40,00){\circle{15}}%
\put( 40,15){\makebox(0,0)[b]{$ $}}%
\put( 40,-15){\makebox(0,0)[t]{$M-2$}}%
\put( 47,00){\line(1,0){26}}%
\put( 80,00){\circle{15}}%
\put( 80,15){\makebox(0,0)[b]{ }}%
\put( 80,-15){\makebox(0,0)[t]{$M-1$}}%
\put( 87,00){\line(1,0){26}}%
\put(127,00){\ldots ~~~~=}%
\put(200,00){\circle{15}}%
\put(200,15){\makebox(0,0)[b]{ }}%
\put(200,-15){\makebox(0,0)[t]{$M-1$}}%
\put(207,00){\line(1,0){26}~~~\dots}%
\put( 35,-5){\line(1, 1){10}}%
\put( 35, 5){\line(1,-1){10}}%
\put(195,-5){\line(1, 1){10}}%
\put(195, 5){\line(1,-1){10}}%
\end{picture}
\end{minipage}
\eea

In the case of $\su(1|1)$ we can now iterate the 
procedure, see diagram \eqref{psu11roots}, and thus derive
indeed the ``free'' Bethe equations \eqref{psu11bethe}
(as first noticed in \cite{CHMS}) written in the Bethe root plane:
\bea\label{psu11betheu}
\left(\frac{u_k+\frac{i}{2}}{u_k-\frac{i}{2}}\right)^L=1
\qquad \qquad
k=1,\ldots,M\, .
\eea
Notice that the locations $u_k$ of the Bethe roots are identical
in the super spin chain and the fermionic spin chain formulation, which
is far from obvious when superficially comparing
\eqref{psu11nested} and \eqref{psu11betheu}.
In fact, this is in line with the intuition that
the central roots $u_k$ roots encode, via $u_k=\half \cot \frac{p_k}{2}$,
the momenta of the physical excitations. These clearly should
not depend on the description.

Turning to the the $\sll(2)$ derivative sector of section
\ref{sl2}, we extract from \cite{BS} that the excitation pattern
of the states \eqref{sl2ops} is 
$M_4=M$, $M_3=M_5=M-1$, $M_2=M_6=M-2$ and $M_1=M_7=0$.
But this means that we can again apply the above reduction 
scheme as symbolized in \eqref{rootreduction}, this time on both
sides of the super Dynkin diagram \eqref{eq:BeautyRoots}.
We thus immediately verify that
\bea\label{sl2embedding}
\begin{minipage}{260pt}
\setlength{\unitlength}{1pt}%
\small\thicklines%
\begin{picture}(260,55)(-10,-30)
\put( 40,00){\circle{15}}%
\put( 40,15){\makebox(0,0)[b]{$ $}}%
\put( 40,-15){\makebox(0,0)[t]{$M-2$}}%
\put( 47,00){\line(1,0){26}}%
\put( 80,00){\circle{15}}%
\put( 80,15){\makebox(0,0)[b]{ }}%
\put( 80,-15){\makebox(0,0)[t]{$M-1$}}%
\put( 87,00){\line(1,0){26}}%
\put(120,00){\circle{15}}%
\put(120,15){\makebox(0,0)[b]{$+1$}}%
\put(120,-15){\makebox(0,0)[t]{$M$}}%
\put(127,00){\line(1,0){26}}%
\put(160,00){\circle{15}}%
\put(160,15){\makebox(0,0)[b]{ }}%
\put(160,-15){\makebox(0,0)[t]{$M-1$}}%
\put(167,00){\line(1,0){26}}%
\put(200,00){\circle{15}}%
\put(200,15){\makebox(0,0)[b]{ }}%
\put(200,-15){\makebox(0,0)[t]{$M-2$}}%
\put(225,-3){$=$}%
\put(260,00){\circle{15}}%
\put(260,15){\makebox(0,0)[b]{$-1$}}%
\put(260,-15){\makebox(0,0)[t]{$M$}}%
\put( 35,-5){\line(1, 1){10}}%
\put( 35, 5){\line(1,-1){10}}%
\put(195,-5){\line(1, 1){10}}%
\put(195, 5){\line(1,-1){10}}%
\end{picture}
\end{minipage}
\eea
This means that now the auxiliary roots ``anti-screen''
the interactions of the roots on the central node,
and we have indeed derived the Bethe equations 
\eqref{sl2betheu} directly from the super chain.
This result opens the interesting possibility to study the
distribution of the auxiliary roots in the 
thermodynamic limit, {\it cf} \cite{BFST}, as it is now fairly
clear how to find their locations given the above discussion.

\section{Three-Loop S-Matrix for the $\cN=4$ Fermionic Sector}
\label{su11section}

Now we would like to illustrate that the principle of non-diffractive
scattering is also very powerful beyond the one-loop level.
Let us apply it to the case of $\su(1|1)$ where the higher loop Bethe 
ansatz is not yet known. However, we do know the Hamiltonian up
to three loops in algorithmic form from the work of \cite{dynamic}. 
Rewriting it in spin chain
form, we find the following two-loop correction\footnote{
The three-loop piece $H_4$ of \cite{dynamic} has been recorded 
(in momentum space form) in \cite{CHMS}; we are refraining
from converting it into spin chain form as it is 
lengthy and little instructive. 
We found it simplest to use Beisert's original code,
and thank him for providing it. 
}
to the XY model Hamiltonian \eqref{xy} of section \ref{psu11}:
\bea\label{2loopxy}
H_2&=&
\sum_{x=1}^L\,\Bigg(
2 (\sigma_x^3-1)
-\sfrac{1}{4}(\sigma_x^3 \sigma_{x+1}^3-1)
+\sfrac{9}{8} (\sigma_x^1 \sigma_{x+1}^1+\sigma_x^2 \sigma_{x+1}^2)
- \\
& &~~~~~~-\sfrac{1}{16} (\sigma_x^1 \sigma_{x+1}^1+\sigma_x^2 \sigma_{x+1}^2)
\sigma_{x+2}^3-
\sfrac{1}{16} \sigma_x^3 (\sigma_{x+1}^1 \sigma_{x+2}^1
+\sigma_{x+1}^2 \sigma_{x+2}^2)-
\nonumber \\
& &~~~~~~~~~~~~~~~~~~~~~~~~~~~~~~-\sfrac{1}{8} 
\sigma_x^1 (1+\sigma_{x+1}^3) \sigma_{x+2}^1
-\sfrac{1}{8} \sigma_x^2 (1+\sigma_{x+1}^3) \sigma_{x+2}^2
\Bigg)\, .\nonumber
\eea
While it
certainly has not yet been rigorously proved that it 
corresponds to an integrable deformation of the one-loop
Hamiltonian, 
spectral studies for small operators in \cite{dynamic} are 
consistent with integrability in that certain tell-tale
degeneracies (the so-called planar pairs argued to be
a hallmark of integrability in \cite{BKS}) indeed reappear.
Further, compelling, evidence will come from the success of the computations
below, as they establish that the spectrum may indeed
be obtained from the principle of factorized scattering.

If one naively extends the approach of section \ref{psu11}
to the higher loop case one quickly finds that the two-body
Bethe ansatz \eqref{oneloopbethe} for the position space wave function 
$\Psi(x_1,x_2)$, defined in \eqref{psu11state}, becomes 
inconsistent. However, we would still expect that the general
form of the Bethe ansatz, namely a superposition of an
in- and outgoing plane wave, is appropriate when the
particles are farther apart than the range of the interaction,
which is, in our case, the considered order of perturbation theory:
\bea
\label{asymptoticbethe}
\Psi(x_1,x_2)\sim e^{i p_1 x_1+i p_2 x_2}+
S(p_2,p_1)~e^{i p_2 x_1+i p_1 x_2}
\qquad {\rm if} \qquad
x_1 \ll x_2
\, .
\eea
This is Sutherland's ``asymptotic'' Bethe ansatz \cite{Sutherland}.
It was used by Inozemtsev to find the Bethe ansatz for the
hyperbolic version of his spin chain \cite{Inoz}, and adapted
in \cite{SS} to diagonalize the three loop dilatation operator
in the bosonic $\su(2)$ sector. 

If the ansatz \eqref{asymptoticbethe} is true,
the Schr\"odinger equation will be satisfied in the asymptotic
region $x_1 \ll x_2$ with the energy value given by the sum over the energies
of the individual partons. One checks that the three-loop dispersion 
law of the $\su(1|1)$ Hamiltonian is the 
same as in the case of the $\su(2)$ chain \cite{SS}, and we thus have
\bea\label{3Ldispersion}
E=\sum_{k=1}^M \Big(
4\, \sin^2\left(\frac{p_k}{2}\right)-
8\, g^2 \sin^4\left(\frac{p_k}{2}\right)+
32\, g^4 \sin^6\left(\frac{p_k}{2}\right)+
\cO(g^6)
\Big)\, ,
\eea
where for the time being $M=2$.
Furthermore, the S-matrix $S(p_1,p_2)$ may be extracted from the 
asymptotics \eqref{asymptoticbethe}, and we expect it to still 
be given by a pure phase
\bea
S(p_1,p_2)=-e^{i \theta(p_1,p_2)}\, .
\eea
How can we find this phase factor? We will need to adapt the method
of the asymptotic Bethe ansatz to the present situation, since, unlike 
in Inozemtsev's case, here we currently know the Hamiltonian 
to three loops only. 
This may be done by modifying the ``fine structure'' of the
wave function close to the collision point,
a technique one might term 
PABA (Perturbative Asymptotic Bethe Ansatz). 
We make the ansatz ($x_1<x_2$), accurate to $\cO(g^4)$:
\bea
\label{higherloopbethe}
\Psi(x_1,x_2)&=&
\Big(1+B_2(p_1,p_2)~g^{2(x_2-x_1)}+B_4(p_1,p_2)~g^{2+2(x_2-x_1)}\Big) 
e^{i p_1 x_1+i p_2 x_2}- \\
& &
-\left(1+C_2(p_1,p_2)~g^{2(x_2-x_1)}+C_4(p_1,p_2)~g^{2+2(x_2-x_1)}\right) 
e^{i p_2 x_1+i p_1 x_2-i \theta(p_1,p_2)}\nonumber \\
& &~~~~+\cO(g^6)\, .
\nonumber
\eea
Note that this form of the wave function is clearly consistent
with the asymptotic ansatz \eqref{asymptoticbethe}.
The intuition behind \eqref{higherloopbethe} is that the number
of powers of the coupling $g^2$ indicates the interaction range
on the lattice. It should be fairly clear how to extend the ansatz
to even higher loop order.

Acting with the three-loop Hamiltonian\footnote{
Note that that the higher loop Hamiltonians are not
uniquely determined and allow for a number of ``gauge parameters'',
{\it cf}~\cite{dynamic}. Some of these will affect the
specific form of the Hamiltonian, as well as the 
wavefunction correction factors, but not the S-matrix (and therefore
the spectrum).
}
on this wavefunction
leads to difference equations similar to (but obviously more
involved than) \eqref{psu11schroedingerO},\eqref{psu11schroedingerII}.
Substituting the perturbative asymptotic Bethe ansatz
\eqref{higherloopbethe} into these equations, one finds,
after somewhat tedious but straightforward computations,
that the Schr\"odinger equation for the
nearest ($x_2=x_1+1$), next-nearest ($x_2=x_1+2$) and
next-to-next-nearest situation ($x_2=x_1+3$) may be satisfied
if we carefully fine-tune the amplitude correction factors
$B_2(p_1,p_2),C_2(p_1,p_2)$ and $B_4(p_1,p_2),C_4(p_1,p_2)$.
At two loops one finds the linear condition
\bea\label{b2c2}
C_2(p_2,p_1)=\sfrac{1}{4}-\sfrac{1}{4}\,e^{2 i p_2-2 i p_1}+
e^{2 i p_2-2 i p_1}\,B_2(p_1,p_2)\, .
\eea
At three loops, the functions $B_2(p_1,p_2)$ and $C_2(p_1,p_2)$
become fully determined, but their detailed form is non-universal
as it depends on various gauge parameters. Furthermore, one
now has a new linear constraint, similar to, but significantly
more complicated than \eqref{b2c2}, which relates 
$B_4(p_1,p_2)$ to $C_4(p_1,p_2)$. We will not display it
as it is not very instructive, and also gauge dependent.
%
%
%
%
The two-body phase shift is then, to three-loop order, derived to be:
\bea
\label{theta}
\theta(p_1, p_2)&=&
4\, g^2 \sin\left(\frac{p_1}{2}\right)\, 
\sin\left(\frac{p_1 - p_2}{2}\right)\,
\sin\left(\frac{p_2}{2}\right)+\\ \nonumber
& & +\, g^4 \sin\left(\frac{p_1}{2}\right) 
\Bigg(\sin\left(\frac{p_1-3\, p_2}{2}\right)
-7\, \sin\left(\frac{p_1-p_2}{2}\right)+\\ \nonumber
& &~~~~~~~~~~~~~~~~~
+ \sin\left(\frac{3\, p_1-3\, p_2}{2}\right)
+\sin\left(\frac{3\, p_1 - p_2}{2}\right)\Bigg)
\sin\left(\frac{p_2}{2}\right)+ \\
& &+\, \cO(g^6)\, \,. \nonumber
\eea
There are many equivalent forms to write this phase shift; 
a further, interesting one is
\bea
\label{thetaalt}
\theta(p_1, p_2)&=&
\bigg(2\, g^2\,
\sin^2\left(\frac{p_1}{2}\right)\,\sin p_2
-2\,g^4\, 
\sin^4\left(\frac{p_1}{2}\right)\,\sin\left(2\,p_2\right)+\\
& &~~~~~~~~~~~~
+8\, g^4\, \sin p_1\, \sin^2\left(\frac{p_1}{2}\right)\,
\sin^2\left(\frac{p_2}{2}\right)+\, \cO(g^6) \bigg)-
\nonumber\\
& &-\big(p_1 \leftrightarrow p_2\big)
\nonumber \, . \nonumber
\eea
%
%
%

The final steps are again identical to the one loop case.
Upon imposing periodic boundary conditions, taking into
account Fermi statistics, 
the Bethe equations are expected to be
\bea
\label{fermibethe}
e^{i p_k L}=
\prod_{\textstyle\atopfrac{j=1}{j\neq k}}^M
e^{i \theta(p_k,p_j)}\, ,
\eea
with $\sum_{k=1}^M p_k=0$.
Now, since we already know their explicit one-loop solution
from section \ref{psu11} we can go further and solve our equations
\eqref{fermibethe} exactly to three-loop order. Taking a logarithm,
we find the fundamental equation ($k=1,\ldots,M$)
\bea
\label{fundamental}
p_k\, L=2\, \pi\, n_k+\sum_{\textstyle\atopfrac{j=1}{j\neq k}}^M\,
\theta(p_k,p_j)\, ,
\eea
which is the higher loop generalization of the free fermion
result \eqref{freefermion}. 
Since we know the scattering phase shift $\theta(p_k,p_j)$
exactly to three loops, {\it cf} \eqref{theta}, we
may solve the fundamental equations recursively to
$\cO(g^4)$ and thus find the loop corrections to the free fermion
momenta \eqref{freefermion}. Finally we substitute the
obtained momenta $p_k$ into the expression 
\eqref{3Ldispersion} for the energy, keeping all terms to
precision $\cO(g^4)$. This procedure then
leads to the following explicit result for the, respectively, one-,
two- and three-loop  anomalous dimensions, where 
$E_{0,2,4}=E_{0,2,4}(L,M,\{n_i\})$:
\bea\label{final}
E_0&=&4 \sum_{k=1}^M \sin^2\left(\frac{\pi n_k}{L}\right)\, ,\\
\nonumber
E_2&=&-8 \sum_{k=1}^M \sin^4\left(\frac{\pi n_k}{L}\right)+\\
& &~~+\frac{16}{L} \sum_{k,j=1}^M \cos\left(\frac{\pi n_k}{L}\right) 
\sin^2\left(\frac{\pi n_k}{L}\right) \sin\left(\frac{\pi n_j}{L}\right)
\sin\left(\frac{\pi (n_k-n_j)}{L}\right)\, ,\nonumber \\
\nonumber
E_4&=&32 \sum_{k=1}^M \sin^6\left(\frac{\pi n_k}{L}\right)+\\
& &~~-\frac{16}{L} \sum_{k,j=1}^M \cos\left(\frac{\pi n_k}{L}\right) 
\sin^2\left(\frac{\pi n_k}{L}\right) \sin\left(\frac{\pi n_j}{L}\right)
\sin\left(\frac{\pi (n_k-n_j)}{L}\right) \times \nonumber \\
& &~~~~~~~~~~~~~~~~~~~\times \left(
5\, \sin^2\left(\frac{\pi n_k}{L}\right)+
\sin^2\left(\frac{\pi n_j}{L}\right)+
\sin^2\left(\frac{\pi (n_k-n_j)}{L}\right) 
\right)\, +\nonumber \\
& &~~+\frac{16}{L^2} \sum_{k,j,m=1}^M
\cos\left(\frac{\pi n_k}{L}\right) 
\sin\left(\frac{\pi n_k}{L}\right) 
\sin\left(\frac{\pi n_j}{L}\right)
\sin\left(\frac{\pi n_m}{L}\right)
 \times \nonumber \\
& &~~~~~~~~~~~~~~~~~~~\times 
\sin\left(\frac{\pi (n_j-n_m)}{L}\right)
\left(\cos\left(\frac{2 \pi n_j}{L}\right)-
\cos\left(\frac{2 \pi (n_k-n_j)}{L}\right) \right)
\nonumber \\
& &~~+\frac{2}{L^2} \sum_{k,j,m=1}^M
\sin\left(\frac{\pi n_k}{L}\right)
\sin\left(\frac{\pi n_m}{L}\right)
\sin\left(\frac{\pi (n_k-n_m)}{L}\right) \times
\nonumber \\
& &~~~~~~~~~~~~~~~~~~\times 
\Bigg(
\sin\left(\frac{2 \pi n_j}{L}\right)+
\sin\left(\frac{2 \pi (n_j-n_k)}{L}\right)+
\sin\left(\frac{2 \pi (n_j+n_k)}{L}\right)-
\nonumber \\
& &~~~~~~~~~~~~~~~~~~~~~~~~~~~~~~~~~~~~~~~~~~
-3\,  \sin\left(\frac{2 \pi (n_j-2 n_k)}{L}\right)
-3\, \sin\left(\frac{4 \pi n_k}{L}\right)
\Bigg)\, ,
\nonumber
\eea
where the relation to the gauge theory scaling dimensions of
the operators \eqref{psu11ops} is given by
\bea\label{Delta3}
\Delta=L+\half\, M+g^2\, E_0+g^4\, E_2+g^6\, E_4+\cO(g^8)\, ,
\quad {\rm with} \quad 
g^2=\frac{\gym^2 N}{8 \pi^2}\, .
\eea

We should stress that these formulas give the {\it explicit, complete} 
three-loop spectrum of planar anomalous dimensions of all
$\cN=4$ operators of the form $\Tr Z^{L-M} \psi^M$. 
The fact that it is possible to find such a result is one
of the amazing consequences of the higher-loop integrability
of the $\cN=4$ gauge theory, as first conjectured in \cite{BKS}. 
They may also be considered as an all-impurity generalization of
the $M=2$ formulas first presented in \cite{BKS} to arbitrary
$M$. (Recall that the spectrum of ``two-impurity'' states agrees in 
all sectors \cite{2imp}). Indeed, one checks that the expressions 
in \eqref{final} reduce, after putting $L=J+1$ and 
$n_1=-n_2:=n$, to eqs.(5.17) and (8.10) of \cite{BKS}.
One may also successfully compare to the spectrum of the first few
lowest states as obtained by direct diagonalization of the
three-loop Hamiltonian for small lengths $L$. In particular,
Beisert worked out the full spectrum of highest weight states
of the sector $\su(2|3)$ up to dimension $\Delta_0=8.5$ 
\cite{dynamic} (see also \cite{thesis}, p.143). 
While none of the $\su(1|1)$ operators are primaries
in $\su(2|3)$, it is straightforward to find the weight
of the corresponding primary operator, and compare
its anomalous dimension with \eqref{final}. We present
the results of this comparison in Table 1. The agreement
is perfect.

One may also check that the formulas \eqref{final} explain 
the numerical results obtained in \cite{CHMS}. Finally, it is 
straightforward to extract the leading $1/J$ correction to 
the BMN limit, for an arbitrary number of ``impurities'',
from \eqref{final}. The result agrees with near-BMN 
string theory at two loops, and disagrees at three,
as was found (numerically) in \cite{CHMS}.

\section{S-Matrices for Quantum Strings at Large Tension}
\label{stringsection}

Let us now turn to the string side and discuss the,
admittedly circumstantial, evidence
that the quantum sigma model might also be described by
elementary excitations living on a circle of 
length\footnote{One of the deep questions relates to the
possible meaning of this ``length'' in the string sigma model.
This is clearly related to the question of the nature of
the elementary excitations on the string side.} 
$L$, and whose scattering is diffractionless. 
That is, we will try to interpret all available
information on the string spectrum in the basic two-component
sectors $\su(2)$, $\su(1|1)$ and $\sll(2)$ in the light
of a (conjectured) factorized S-matrix. We then expect
that the underlying equation describing this spectrum
should be of the type of a {\it fundamental} equation
\bea
\label{fundamental2}
p_k\, L=2\, \pi\, n_k+\sum_{\textstyle\atopfrac{j=1}{j\neq k}}^M\,
\theta(p_k,p_j)\, ,
\eea
with $\sum p_k=0$, obtained by taking a logarithm on both sides of the
Bethe equation. Here the scattering phase shift 
$\theta(p_k,p_j)$ is related to the S-matrix as
$\theta(p_k,p_j)=- i \log \left( \pm S(p_k,p_j) \right)$ 
(the upper sign is for bosons and the lower for fermions),
$p_k$ are the momenta of individual
excitations, and the $n_k$ are quantum numbers.
The total momentum $P=Q_1$, the total
energy $E=Q_2$, as well as all other higher charges $Q_r$ are
then expected to be given 
as linear sums over local dispersion laws 
$q_r(p_k)$:
\bea\label{charges}
Q_r=\sum_{k=1}^M\,q_r(p_k)\, .
\eea

\paragraph{Finite Length Dispersion Laws}
The first charge $q_1(p_k)$ is the momentum $p_k$:
\bea\label{q1}
q_1(p_k)=p_k\, .
\eea
In \cite{BDS} we proposed for the second charge $q_2(p_k)$, 
i.e.~for the energy per excitation, the following long-range 
dispersion law
\bea\label{q2}
q_2(p_k)=
\frac{1}{g^2}\left(\sqrt{1+8\,g^2\,\sin^2\left(\frac{p_k}{2}\right)}\, -\, 1
\right)\, ,
\eea
with
\bea
g^2=\frac{\lambda}{8 \pi^2}=\frac{\gym^2 N}{8 \pi^2}\, ,
\eea
where $\sqrt{\lambda}$ is the string tension.
This is simply a lattice version of the famous BMN energy
formula \cite{BMN}; in fact, one may show that this is the
{\it only} possible lattice discretization of the BMN expression
\cite{BKS,RT}.

Furthermore, a study of the dispersion laws for the higher charges in
the $\su(2)$ sector, up to five loops, suggests \cite{BDS} 
that these are likely given by the expressions
\bea \label{qr}
q_r(p_k)=\frac{2\sin\left(\sfrac{r-1}{2}\,p_k\right)}{r-1}
\left(\frac{\sqrt{1+8\,g^2\,\sin^2\left(\frac{p_k}{2}\right)}\, -\, 1
}{2\,g^2\,\sin\left(\frac{p_k}{2}\right)} \right)^{r-1} \, .
\eea
We expect \eqref{qr} to hold for the elementary excitations in all sectors.

\paragraph{BMN Limit}

In this limit \cite{BMN} one takes $L\rightarrow\infty$ while keeping
$M=2,3,\ldots$ small. This is a dilute gas approximation, where,
except for level matching, i.e.~momentum conservation, the
excitations, in both gauge and string theory,
do not feel each others presence. Put differently \cite{MZ}, 
we have
\bea
\theta(p_k,p_j) \simeq 0\, ,
\eea
in \eqref{fundamental2}, i.e.~there is {\it no scattering} at all!
%
%
The fundamental equation \eqref{fundamental2} then simply
leads to the quantization law of free non-interactive particles
on a circle: $p_k~L=2 \pi n_k$. Substituting this solution of
the ``Bethe equations'' into the dispersion law \eqref{q2} one then 
immediately recovers, via $\Delta=1+g^2~Q_2$ and using 
$L \simeq J$, the BMN formula
\bea\label{bmn}
\Delta=J+\sum_{k=1}^M\,\sqrt{1+\lambda'\,n_k^2}+
\frac{\delta \Delta}{J}+
\cO(\frac{1}{J^2})\, 
\qquad {\rm with} \qquad
\lambda'=\frac{\gym^2 N}{J^2}\, .
\eea
where we have also already introduced the notation $\delta \Delta$
for the leading order $\cO(1/J)$ correction to the energy, see below.

As was first suggested in \cite{BDS,AFS}, this simple picture leads to
an interesting interpretation of the {\it discrepancies} between
gauge and string theory which show up in the 
near-BMN limit \cite{Call1} and the Frolov-Tseytlin limit \cite{SS},
but, strikingly, apparently {\it not in the strict BMN limit}:
The discrepancy might be due to a change in the S-matrix (and thus
the phase shift $\theta(p_k,p_j)$) as we go from weak to strong
coupling\footnote{It will also be apparent that both limits probe
the phase function and thus the S-matrix to leading order in 
small values for the individual excitation momenta. These are, unlike
in finite length gauge theory, of order $\cO(1/L)$ in both
the near-BMN and FT situation. There are indeed indications that
the discrepancies might worsen when higher corrections
(i.e. $1/J^2$ corrections to BMN and $1/J$ corrections to FT)
are considered \cite{LZ,FPT}. This is entirely consistent
with our suspicion that the S-matrix is to be blamed. It 
presumably changes as one goes from weak to strong coupling.}. 
In this picture, the local dispersion laws should hold
for arbitrary values of the coupling constant. Let us next
discuss these refined limits, where the interactive but integrable
nature of the CFT/AdS system begins to emerge.

\paragraph{Near-BMN Limit} 

Here one is interested in the first $\cO(1/J)$ correction
to the BMN expression \eqref{bmn}. One then expects 
corrections\footnote{There are some subtleties, not important
for the present discussions, when some of the
mode numbers coincide \cite{MZ,AFS}. 
It was explained in detail in \cite{AFS} how to deal with them,
and the procedure immediately applies to the more general 
setting discussed here.} to the free particle motion:
\bea\label{nbmnfundamental}
p_k=\frac{2\,\pi}{L}\,n_k+\frac{\delta p_k}{L^2}+\cO(\frac{1}{L^3})\, .
\eea
What happens is that, due to finite volume effects, the
dilute gas approximation breaks down and leading order scattering
phase shifts have to be taken into account.
If the S-matrix is known, the leading momentum shifts are immediately
found from the fundamental equation \eqref{fundamental2}:
\bea\label{nbmnshift}
\delta p_k=L\, \sum_{\textstyle\atopfrac{j=1}{j\neq k}}^M\,
\theta(\sfrac{2 \pi}{L}n_k,\sfrac{2 \pi}{L}n_j)\, ,
\eea
where it is understood that we only keep terms to leading order in 
$1/L$ in the expression 
$\theta(\sfrac{2 \pi}{L}n_k,\sfrac{2 \pi}{L}n_j)$.
Some care has to be taken as the difference between the length
$L$ of the system and the R-charge $J$ begins to matter at
this order. For the three basic two-component sectors
one has $L=J+\nu M$, where, $\nu=1,\half,0$ for, respectively
$\su(2)$, $\su(1|1)$ and $\sll(2)$.  
Expanding the dispersion law \eqref{q2} to this order, we find
the following near-BMN energy shift
\bea\label{nbmneng}
\delta \Delta=
\lambda'\,\sum_{k=1}^M\, \frac{n_k}{\sqrt{1+\lambda'\,n_k^2}}\,
\left(\frac{\delta p_k}{2 \pi}-\nu\,M\,n_k\right)\, .
\eea
We may then combine \eqref{nbmnshift} and \eqref{nbmneng} into the
final formula
\bea\label{nbmneng2}
\delta \Delta=
\lambda'\,\sum_{\textstyle\atopfrac{k,j=1}{j\neq k}}^M\,
\frac{n_k}{\sqrt{1+\lambda'\,n_k^2}}\,
\left(\frac{L}{2 \pi}\,\theta(\sfrac{2 \pi}{L}n_k,\sfrac{2 \pi}{L}n_j)\,
+\nu\,(n_j-n_k)\right)\, .
\eea
Expressions of this structure are indeed found in recent
studies of the exact multi-oscillator quantization of strings 
in the near-plane wave geometry \cite{Call2,McLS}. 
Below, {\it cf} section \ref{psu11string}, we will use this connection 
in reverse, and obtain important information on the string 
S-matrices from the results of \cite{Call2,McLS}.

\paragraph{Frolov-Tseytlin Limit}

In the thermodynamic limit, around the ferromagnetic vacuum
corresponding to BPS states, the fundamental equation \eqref{fundamental2}
is expected to turn into an integral equation after the
rescaling $p_k L \rightarrow p_k$ 
\cite{BMSZ,BFST,SS,BDS}
\bea\label{thermofundamental}
p(\varphi)=2\,\pi\,n_{\nu}
+\pint_{\contourgauge} d\varphi'\,\rho(\varphi')\,\theta(\varphi,\varphi')\, 
\quad {\rm with} \quad
\varphi \in \contourgauge_\nu\, .
\eea
This limit was originally considered in order to
to compare gauge and string theory,
following an inspiring proposal by Frolov and Tseytlin \cite{FT}.
Note that \eqref{thermofundamental} is analogous to the
equations \eqref{nbmnfundamental},\eqref{nbmnshift} of the
near-BMN situation in that we expand the phase shift to the same
leading order in $\cO(1/L)$. The only difference is that
a large number $M=\cO(L)$ of excitations have degenerate mode numbers, 
which requires to work out how their mutual degeneracy is lifted. (This
is analogous to the refined near-BMN situation where one also
considers coinciding mode numbers.)
In \eqref{thermofundamental} 
$\varphi$ is a convenient spectral parameter which leads
to a simple form of the momentum $p=p(\varphi)$, all other
charges $q_r(\varphi):=q_r(p(\varphi))$ as well as the 
phase shift $\theta(\varphi,\varphi'):=\theta(p(\varphi),p(\varphi'))$.
The discrete ``Bethe roots'' $\varphi_k$ are expected to densely
assemble on a union of smooth contours 
$\contourgauge=\contourgauge_1 \cup \contourgauge_2 \ldots$, such
that all roots on each component $\contourgauge_\nu$
carry the same quantum number $n_\nu$. Note that
this excludes the existence of this particular
type of thermodynamic limit in the fermionic sector, as Fermi statistics
does not allow coinciding quantum numbers.
$\rho(\varphi)$ is the distribution density of the Bethe roots
$\varphi$, with support on $\contourgauge$ in the complex plane.
It is normalized as
\bea\label{density}
\frac{M}{L}=\int_{\contourgauge} d \varphi\,\rho(\varphi)\, ,
\eea
i.e.~it counts the number of excitations in units of the length
of the circle (however, {\it cf}~footnote at the beginning of this
chapter).
The charges \eqref{charges} are then given in the thermodynamic limit, 
after rescaling
$q_r \, L^r\, \rightarrow q_r$ and $Q_r \, L^r\, \rightarrow Q_r$,
by
\bea\label{thermocharges}
Q_r=\int_{\contourgauge} d \varphi\,\rho(\varphi)\,q_r(\varphi)\, .
\eea
Equations very similar to \eqref{thermofundamental},\eqref{density},
\eqref{thermocharges} are indeed found in recent
studies of semiclassical strings moving on the sphere \cite{KMMZ},
or on AdS \cite{KZ}. In \cite{AFS} we used this connection 
in reverse, and obtained important information on the string 
S-matrix of the $\su(2)$ sector from the results of \cite{KMMZ}.
The salient points will be briefly reviewed in the next section 
\ref{su2string}. Later we show in section \ref{sl2string} that
similar results may be obtained for the $\sll(2)$ sector,
where equations resembling Bethe equations
have also become available \cite{KZ}.

\subsection{The $\su(2)$ Bosonic Sector}
\label{su2string}

In \cite{KMMZ} an equation describing the semiclassical spectrum
of strings moving with two large angular momenta on the
five-sphere was obtained. We would expect the corresponding states
to be related to gauge theory operators in the $\su(2)$ sector
\cite{FT}. This equation reads
\bea\label{su2stringbethe}
\frac{x+2\, \omega^2\,E\,x}{x^2-\omega^2}=
2\, \pi\, n_\nu+
2\, \pint_{\cC} dx'\,\frac{\sigma(x')}{x-x'}
\qquad \mbox{with} \qquad
x \in \cC_\nu \, ,
\eea
with the energy\footnote{In much of the literature on 
the sigma model, including \cite{AFS}, the total string energy
is $1+2 \omega^2 E$. Here we prefer to continue to use spin chain
terminology where the energy $E$ is the eigenvalue of the
Hamiltonian, i.e.~the anomalous part of the dilatation operator
divided by $g^2$.
We will also continue to use the same symbols in the finite $L$ and 
the (rescaled) large $L$ cases, as it should always be clear from
the context what is meant.}
$E$ and the momentum $P$ given by
\bea\label{momeng}
E=\int_{\cC} dx\, \frac{\sigma(x)}{x^2}
\qquad {\rm and} \qquad
P=\int_{\cC} dx\, \frac{\sigma(x)}{x}\, ,
\eea
where $P$ is quantized by an integer $m$: $P=2 \pi m$, and the
coupling constant $\omega$ appropriate for the thermodynamic limit is
\bea\label{omega}
\omega^2=\frac{g^2}{2 L^2}=
\frac{\gym^2 N}{16 \pi^2 L^2}\, .
\eea
(For comparison, note that $\omega^2$ is denoted
by $T=\omega^2$ in \cite{KMMZ,KZ}.) 

The spectral equation \eqref{su2stringbethe} superficially resembles
the fundamental equation \eqref{thermofundamental} of the scattering
approach, and one might be tempted to identify
the spectral parameter $x$ with $\varphi$ in \eqref{thermofundamental}. 
However, the fact that an extensive quantity,
namely the {\it energy} $E$, appears on the left hand side of 
\eqref{su2stringbethe} prevents us from interpreting the latter
as a parton momentum $p(x)$. A related problem is that
the function $\sigma(x)$ is found in \cite{KMMZ} to be normalized as
\bea\label{norm}
\frac{M}{L}=
\int_{\cC} dx\, \sigma(x)\, \left(1-\frac{\omega^2}{x^2}\right)\, ,
\eea
which does not allow for an interpretation of $\sigma(x)$ as
an excitation density as in \eqref{density}. However,
it was shown in \cite{BDS} that both problems may be solved ``in one go'' if
we change spectral parameters from $x$ to $\varphi$
according to 
\bea\label{map}
\varphi=x+\frac{\omega^2}{x}\, ,
\qquad {\rm with} \qquad
\rho(\varphi):=\sigma(x)\, .
\eea
For one, the quantity $\rho(\varphi)$ in \eqref{map} is
then indeed normalized as in \eqref{density}, and secondly,
the spectral equation \eqref{su2stringbethe} may now
be rewritten in the form of a fundamental scattering equation 
\eqref{thermofundamental} with the following dependence of the
excitation momenta on the spectral parameter: 
\bea\label{p}
p(\varphi)=\frac{1}{\sqrt{\varphi^2-4\,\omega^2}}\, .
\eea
%
%
The two-body phase shift is then found to be
\bea\label{su2shift}
\theta_{\su(2)}\upper{string}(\varphi,\varphi')
=2\, \theta_0(\varphi,\varphi')+
2\, \sum_{r=2}^\infty\,\theta_r(\varphi,\varphi')\, ,
\eea
where
\bea\label{hilbert}
\theta_0(\varphi,\varphi')=\frac{1}{\varphi-\varphi'}\, ,
\eea
with an infinite number of additional contributions 
$2\,\theta_r$ which may be suggestively written as \cite{AFS}
\bea\label{thetaphi}
\theta_r(\varphi,\varphi')=
\omega^{2 r}\,\big(q_r(\varphi)\, q_{r+1}(\varphi')
-q_{r+1}(\varphi)\, q_r(\varphi')\big)\, .
\eea
Here the $q_r$ are found to be given by the expressions 
\bea
\label{qrphi}
q_r (\v) = \frac{1}{\sqrt{\varphi^2-4\o^2}}\frac{1}
{\left(\frac{1}{2}\v+\frac{1}{2}\sqrt{\varphi^2-4\o^2}\right)^{r-1}}\, .
\eea
It so turns out that these formulas are {\it identical} to
the thermodynamic limit of our lattice dispersion laws 
\eqref{qr} when expressed, via \eqref{p}, as functions of the 
spectral parameter $\varphi$. This is fascinating, as they
simply follow from rewriting the semiclassical equations
of \cite{KMMZ} in a form that unveils the underlying
scattering processes! 

What is more, recovering the thermodynamic fundamental scattering 
equation from classical string theory suggests a very natural way
to rediscretize it and guess the large tension, small momentum
(i.e.~large length $L$) S-matrix 
of the quantum sigma model \cite{AFS}. Here we recall that the
proposed \cite{BDS} long-range S-matrix of the weak-coupling gauge theory
in the $\su(2)$ sector is 
\bea
\label{bds}
S_{\su(2)}\upper{gauge}(p_k,p_j)&=&
\frac{\varphi(p_k)-\varphi(p_j)+i}{\varphi(p_k)-\varphi(p_j)-i}\, .
\eea
Its logarithm yields, in the thermodynamic limit, the leading contribution 
$\theta_0$, see \eqref{hilbert}, to the string theory
two-body scattering kernel. 
The all-order expression for the lattice phase function $\varphi(p_k)$ 
is conjectured \cite{BDS} to be 
\bea
\label{pf} 
\varphi(p_k)=\half\cot\left(\frac{p_k}{2}\right)
\sqrt{1+8\,g^2\,\sin^2\left(\frac{p_k}{2}\right)} \, ,
\eea
and yields in the thermodynamic limit, upon inversion, \eqref{p}.
It is then very natural to include the additional scattering
terms $\theta_r$ of \eqref{thetaphi} as a ``dressing factor''
to the ``bare'' S-matrix \eqref{bds} and write down an ansatz
for the string S-matrix\cite{AFS}:
\bea
\label{Ssu2}
S_{\su(2)}\upper{string}(p_k,p_j)&\simeq&
 \frac{\varphi(p_k)-\varphi(p_j)+i}{\varphi(p_k)-\varphi(p_j)-i}\,
\prod_{r=2}^\infty \,
e^{
2 i \theta_r(p_k,p_j)}
\, ,
\eea
while replacing (1), via \eqref{omega}, the continuum coupling $\omega^2$ by
the lattice coupling $g^2$ and (2) the continuum dispersion laws
\eqref{qrphi} by the lattice laws \eqref{qr}: 
\bea\label{thetap}
\theta_r(p_k,p_j)=
\Big(\frac{g^2}{2}\Big)^r\,
\big(q_r(p_k)\, q_{r+1}(p_j)
-q_{r+1}(p_k)\, q_r(p_j)\big)\, .
\eea
One now checks, with the help of \eqref{nbmneng2}, agreement with the
near-BMN multi-impurity results of \cite{Call2,McLS}, as was first
done in \cite{AFS}.

\subsection{The $\su(1|1)$ Fermionic Sector}
\label{psu11string}
In the fermionic sector no ``ferromagnetic'' thermodynamic limit 
similar to the one discussed in the last section exists. Due to Fermi 
statistics the mode numbers of all excitations are distinct. This prevents
the Bethe roots from condensing onto smooth cuts in the complex spectral 
parameter plane, where the mode number has to stay constant along each 
contour. Luckily there is an alternative way to deduce information on
the string S-matrix of this sector. In a beautiful paper Callan,
McLoughlin and Swanson \cite{Call2} studied the spectrum of {\it three} 
elementary string excitations in the near-BMN limit. It so turns out that 
this is precisely what we need! In section \ref{oneloop} we argued
that, given integrability, it suffices to carefully solve the
two-body problem in order to deduce the S-matrix. It is however
important that the two excitations are ``off-shell'', i.e.~they
need to be capable of carrying arbitrary momenta. Now, since the level 
matching condition 
in string theory enforces the total momentum conservation of all 
excitations, the solution of the two-impurity problem \cite{Call1}
does not contain enough information as the absolute values of
the two associated momenta are necessarily equal. This is no
longer the case if we decompose the three-body scattering
into a series of two-body processes. And indeed, the principle
of factorized scattering then immediately yields the solution of
the M-body problem, as was 
intuitively and correctly understood in a nice follow-up paper by
McLoughlin and Swanson \cite{McLS}. 
Rewritten in our present notations, they find 
({\it cf} eq.(3.30) in \cite{McLS}) the following energy shift 
for $M$ excitations in the $\su(1|1)$ sector:
\bea\label{mcls-su11}
\delta \Delta=-
\frac{\lambda'}{4}\,\sum_{\textstyle\atopfrac{k,j=1}{j\neq k}}^M\,
\left(\frac{n_k^2+n_j^2+2\,\lambda'\,n_k^2\,n_j^2}
{\sqrt{1+\lambda'\,n_k^2}\sqrt{1+\lambda'\,n_j^2}}\,-
2\,n_k\,n_j
\right)\, .
\eea
This double sum may be rearranged as
\bea
\delta \Delta=
\lambda'\,\sum_{\textstyle\atopfrac{k,j=1}{j\neq k}}^M\,
&&\frac{n_k}{\sqrt{1+\lambda'\,n_k^2}} \times \\ \nonumber
&&\times
\Bigg(\half\,n_j\,\left(\sqrt{1+\lambda'\,n_k^2}-1\right)-
\half\,n_k\,\left(\sqrt{1+\lambda'\,n_j^2}-1\right)
+\half\,(n_j-n_k)\Bigg)\, 
\eea
By comparing to our general formula \eqref{nbmneng2} 
(here $\nu=\half$ as appropriate for the fermionic sector),
we may then
read of the phase shift, accurate to leading order in 
small $p_k$, $p_j$, as
\bea
-\theta_1(p_k,p_j)\simeq-\frac{g^2}{2}
\big(p_k\, q_{2}(p_j)-q_{2}(p_k)\, p_j\big)
\, ,
\eea
where we have used \eqref{q2} (for small $p_k$) and employed the notation
$\theta_r(p_k,p_j)$, introduced in \eqref{thetap}, with $r=1$.
This then leads to the following simple, approximate S-matrix
encoding the near-BMN string physics in the fermi sector:
\bea
\label{Spsu11}
S_{\su(1|1)}\upper{string}(p_k,p_j)&\simeq&
-e^{
-i \theta_1(p_k,p_j)}\, .
\eea

\subsection{The $\sll(2)$ Derivative Sector}
\label{sl2string}

In the case of the third basic two-component sector $\sll(2)$ we
have a choice for extracting the S-matrix from known string theory results.
We could either proceed as in the last section \ref{psu11string}, using
again \eqref{nbmneng2} in conjunction with 
the multi-impurity near-BMN results of
\cite{Call2,McLS}. Alternatively we can apply once more
the logic of section \ref{su2string} and test them on 
the semiclassical string sigma model results of Kazakov and Zarembo 
which have recently become available \cite{KZ}. 
Interestingly, both procedures lead to the same final result,
showing once more the close connection between the leading curvature 
corrections to quantum strings in a plane wave geometry, and semiclassical 
strings on the curved geometry $AdS_5 \times S^5$. Let us proceed
in the second fashion, which will serve as a nice check on the ideas
presented in section \ref{su2string}.

For semiclassical strings rotating with one large angular momentum on the
five-sphere and one large spin on AdS, the bootstrap equation
derived in \cite{KZ} from the monodromy matrix of the
classically integrable string sigma model reads
\bea\label{sl2stringbethe}
\frac{x-2\, \omega^2\, P}{x^2-\omega^2}=
2\, \pi\, n_\nu-
2\, \pint_{\cC} dx'\,\frac{\sigma(x')}{x-x'}
\qquad \mbox{with} \qquad
x \in \cC_\nu \, ,
\eea
where the function $\sigma(x)$ is normalized {\it exactly} as
in the S$^5$ case according to formula \eqref{norm} (now
$M$ is the spin quantum number on AdS$_5$, whereas previously
$M$ was one of the two angular momenta on S$^5$, see also
\cite{BFST}). The equation \eqref{sl2stringbethe}
is manifestly very similar to the corresponding equation
\eqref{su2stringbethe} of the $\su(2)$ sector. Again an
extensive quantity appears on the left side
of this equation, preventing its naive interpretation as 
a fundamental scattering equation: This time it is not the energy 
$E$ but the {\it total momentum} $P$, see \eqref{momeng}.
However, since the normalization conditions \eqref{norm} are
identical in the two cases, it is natural to once more apply the
same change of spectral parameter \eqref{map}. This converts
\eqref{sl2stringbethe} to the form of a fundamental scattering
equation, see \eqref{thermofundamental}
\bea\label{sl2fundamental}
p(\varphi)=2\,\pi\,n_{\nu}
-\pint_{\contourgauge} d\varphi'\,\rho(\varphi')\,
\left(\frac{2}{\varphi-\varphi'}+
2\, \sum_{r=1}^\infty\,\theta_r(\varphi,\varphi')\right)\,
\eea
where $\varphi \in \contourgauge_\nu$, the momentum is parametrized
as before through \eqref{p}, the density
$\rho(\varphi)$ is consistently normalized as in \eqref{density},
and the phase shifts $\theta_r(\varphi,\varphi')$ are given
in \eqref{thetaphi}.
We may then read off the continuum two-body scattering phase shift 
of the $\sll(2)$ sector as
\bea\label{sl2shift}
\theta_{\sll(2)}\upper{string}(\varphi,\varphi')
=-2\, \theta_0(\varphi,\varphi')-
2\, \sum_{r=1}^\infty\,\theta_r(\varphi,\varphi')\, ,
\eea
which differs from the one of the $\su(2)$ sector \eqref{su2shift}
by (1) an overall minus sign and (2) the fact that the sum starts
at $r=1$ instead of $r=2$.

As a consistency check we may verify that the
near-BMN physics is properly reproduced. Replacing the shift
\eqref{sl2shift} by its discrete, small $p_k$ (i.e.~large $L$)
version in the, by now, familiar fashion, we find
\bea\label{nbmnsl2}
\theta(p_k,p_j)\simeq
-\frac{2}{\varphi(p_k)-\varphi(p_j)}-
2\, \sum_{r=1}^\infty\,\theta_r(p_k,p_j)\, ,
\eea
where the lattice expressions \eqref{pf} and \eqref{thetap} should
be used (replacing momenta by their leading $\cO(1/L)$ approximations).
If we now insert \eqref{nbmnsl2} into the formula \eqref{nbmneng2} 
for the energy shift (with $\nu=0$ for the $\sll(2)$ sector) we
reproduce, after a short calculation, indeed the $\sll(2)$ 
multi-impurity result of \cite{Call2,McLS}
\bea\label{mcls-sl2}
\delta \Delta=
\frac{\lambda'}{2}\,\sum_{\textstyle\atopfrac{k,j=1}{j\neq k}}^M\,
\left(\frac{n_k\,n_j-\,\lambda'\,n_k^2\,n_j^2}
{\sqrt{1+\lambda'\,n_k^2}\sqrt{1+\lambda'\,n_j^2}}\,+
\,n_k\,n_j
\right)\, .
\eea
{\it cf} equation (3.24) in \cite{McLS}. The case of coinciding
mode numbers is dealt with exactly as in \cite{AFS}, and immediately
reproduces equation (3.23) of \cite{McLS}.

Finally we may write the (large tension, small momentum)
S-matrix for the $\sll(2)$ sector suggested by the phase shift 
\eqref{nbmnsl2}:
\bea
\label{Ssl2}
S_{\sll(2)}\upper{string}(p_k,p_j)&\simeq&
 \frac{\varphi(p_k)-\varphi(p_j)-i}{\varphi(p_k)-\varphi(p_j)+i}\,
\prod_{r=1}^\infty \,
e^{
-2 i \theta_r(p_k,p_j)}\, .
\eea
Thus the approximate string S-matrix in the $\sll(2)$
sector differs from the one of the $\su(2)$ sector 
\eqref{Ssu2} of \cite{AFS} by (1) complex
conjugating the explicit factors of $i \rightarrow -i$ and (2) by
one additional factor with $r=1$.

\section{Three-Loop S-Matrix for the $\cN=4$ Derivative Sector}

Observe the following striking relationship between the conjectured
large string tension, small momentum S-matrices 
\eqref{Ssu2},\eqref{Spsu11},\eqref{Ssl2}
of the three basic two-component sectors:
\bea\label{grouptheory}
S_{\sll(2)}=S_{\su(1|1)}\, S^{-1}_{\su(2)}\, S_{\su(1|1)}\, .
\eea
The simplicity of the relation \eqref{grouptheory} suggests
a purely group-theoretical explanation.
In fact, it appears to result from an ``inversion'' of
the Dynkin diagram \eqref{eq:BeautyRoots} where we now
place the representation $-1$ onto the central node; 
i.e.~we exchange the AdS and S$^5$ sectors.

Since the relationship \eqref{grouptheory} seems to be based on 
symmetry alone it is very reasonable to expect that it
also holds at weak coupling, i.e.~in the gauge theory. We are thus
led to the following Bethe ansatz
for the weak-coupling $\sll(2)$ sector at finite $L=J$ and finite $M=S$
(with $k=1,\ldots,M$):
\bea\label{wild}
e^{i p_k L}=
\prod_{\textstyle\atopfrac{j=1}{j\neq k}}^M
\Bigg(
\frac{\varphi(p_k)-\varphi(p_j)-i}{\varphi(p_k)-\varphi(p_j)+i}\,
e^{2 i \theta(p_k,p_j)}\Bigg)\, ,
\eea
where $\varphi(p)$ is given in \eqref{pf} and 
$\theta(p_k,p_j)$ is currently known to three-loop precision, 
{\it cf} \eqref{theta} or \eqref{thetaalt}.

A first, non-trivial test of this ansatz may be performed immediately.
We know that all two-impurity states must agree for arbitrary,
finite R-charge $J$ \cite{2imp}. Let us compare the $M=2$ Bethe ans\"atze
for the $\su(2)$ sectors \cite{SS,BDS} and the
$\su(1|1)$ sectors ({\it cf} section \ref{psu11}):
\bea\label{twoimp}
\su(2):
\quad
e^{i p (J+2)}=
\frac{\varphi(p)+\frac{i}{2}}{\varphi(p)-\frac{i}{2}}
\qquad \qquad
\su(1|1):
\quad
e^{i p (J+1)}=e^{i \theta(p,-p)}\, ,
\eea
from which we conclude that we must have to all orders in the 
coupling constant $g$ 
\bea\label{magic}
e^{i \theta(p,-p)}=e^{-i p}\,
\frac{\varphi(p)+\frac{i}{2}}{\varphi(p)-\frac{i}{2}}\, .
\eea
The reason is that we have the same dispersion relation in all
sectors, and the momenta $p_1:=p$, $p_2:=-p$ for two impurities
must therefore agree in different sectors. 
We easily check that the three-loop
phase shift obtained in section \ref{psu11} satisfies \eqref{magic}
to $\cO(g^4)$.
One now immediately verifies that the ansatz \eqref{wild} 
for the $\sll(2)$ sector, where the length is $L=J$, results
in a Bethe equation for $p$ which is entirely equivalent to
\eqref{twoimp}:
\bea\label{wild2imp}
\sll(2):
\quad
e^{i p J}=
\frac{\varphi(p)-\frac{i}{2}}{\varphi(p)+\frac{i}{2}}\,
e^{2 i \theta(p,-p)}\, .
\eea

Now, naively we would expect the Bethe ansatz \eqref{wild} to
be {\it asymptotic} only, that is we may a priori not expect that it
properly diagonalizes operators which are shorter than the range of 
interaction. These are precisely the twist-two operators at
two and three loops, and the twist three operators at three loops. 
However, the consistency check just presented actually yields 
a further, crucial hint. We notice that our $\sll(2)$ ansatz
also works at least up to three loops for the $M=2$ states with 
$J=2$ and $J=3$, {\it cf} \eqref{wild2imp}. But these are, respectively,
length two and length three operators. This observation leads
to the expectation that in fact {\it all} $\sll(2)$ operators
are diagonalized by the Bethe ansatz \eqref{wild}!
This allows us to test our ansatz since, luckily, the
three-loop anomalous dimensions of twist-two operators in
the $\cN=4$ gauge theory have recently
become available. Kotikov, Lipatov, Onishchenko and Velizhanin 
\cite{KLOV} were able to extract them, under some unproven but 
astute assumptions, from an impressive, rigorous
computation of three-loop anomalous dimensions in QCD 
by Moch, Vermaseren and Vogt \cite{MVV}.

We have not worked out the explicit three-loop spectrum of twist
two operators from \eqref{wild} for arbitrary even
AdS spin $M$
(in the $\sll(2)$ sector $S=M$ has to be even for twist-two operators),
but certainly 
expect that this will reproduce the general formulas derived 
in \cite{KLOV} ({\it cf} equations (10),(11),(12) of that paper,
where the notation $j=M$ is used), which read:
\bea
\label{klov}
E_0(M) &=& 4\, \Sigma_1\, ,  \\
E_2(M) &=&-4\,\Big( \Sigma_{3} + \Sigma_{-3}  -
2\,\Sigma_{-2,1} + 2\,\Sigma_1\,\big(\Sigma_{2} + \Sigma_{-2}\big) \Big)\, ,  
\nonumber \\
E_4(M) &=& -8 \Big( 2\,\Sigma_{-3}\,\Sigma_2 -\Sigma_5 -
2\,\Sigma_{-2}\,\Sigma_3 - 3\,\Sigma_{-5}  +24\,\Sigma_{-2,1,1,1}\nonumber\\
&&~~~~~~+ 
6\,\big(\Sigma_{-4,1} + \Sigma_{-3,2} + \Sigma_{-2,3}\big)
- 12\,\big(\Sigma_{-3,1,1} + \Sigma_{-2,1,2} + \Sigma_{-2,2,1}\big)\nonumber \\
&&~~~~~~-
\big(\Sigma_2 + 2\,\Sigma_1^2\big) 
\big( 3 \,\Sigma_{-3} + \Sigma_3 - 2\, \Sigma_{-2,1}\big)
- \Sigma_1\,\big(8\,\Sigma_{-4} + \Sigma_{-2}^2\nonumber \\
&&~~~~~~+ 
4\,\Sigma_2\,\Sigma_{-2} +
2\,\Sigma_2^2 + 3\,\Sigma_4 - 12\, \Sigma_{-3,1} - 10\, \Sigma_{-2,2} 
+ 16\, \Sigma_{-2,1,1}\big)
\Big)\, , \nonumber
\eea
and
the harmonic sums $ \Sigma_{\pm a,b,c,\cdots} :=\Sigma_{\pm a,b,c,\cdots}(M)$
are defined recursively ($a,b,c > 0$)
%
\bea\label{harmonic}
%
\hspace*{-1cm} \Sigma_{\pm a}(M)~=~ \sum^M_{m=1} \frac{(\pm 1)^m}{m^a},
\qquad
\Sigma_{\pm a,b,c,\cdots}(M)~=~ \sum^M_{m=1} \frac{(\pm 1)^m}{m^a}\,
\Sigma_{b,c,\cdots}(m)\, .  \nonumber 
\eea
We checked explicitly that our ansatz \eqref{wild} reproduces the result
predicted in \eqref{klov} in the cases $M=2,4,6,8$; see also
Table 2.

We would however like to stress that the ansatz \eqref{wild}
is expected to also properly reproduce the three-loop anomalous
dimensions for operators of arbitrary twist and spin.
As far as we know no non-trivial twist-three two-loop, let alone three-loop,
anomalous dimensions seem to be known in $\cN=4$ gauge
theory from a field theory computation to date.
For example, for the simplest twist-three field not in
the two-impurity supermultiplet, namely 
$\Tr D^3 Z^3 + \ldots$, our ansatz predicts a paired state with
three-loop energy
\bea\label{twistthree}
E=\frac{15}{2}-\frac{225}{16}\,g^2+\frac{3195}{64}\,g^4\, 
+ \cO(g^6)\, .
\eea
It would be exciting if this prediction could be checked
by a traditional Feynman diagram computation\footnote{
After the completion of this manuscript we were informed by
B.~Eden that the two-loop part of the prediction \eqref{twistthree}
may indeed be confirmed by a full-fledged field theory 
computation \cite{confirm}.} 

Finally we may answer a question raised in the paper
\cite{KZ} about whether the discrepancies between string and
gauge theory might already appear in the $\sll(2)$ sector
at {\it two} instead of at three loops. Comparing the S-matrix 
entering  the r.h.s.~of \eqref{wild} ({\it cf} also \eqref{proposal13})
and the string S-matrix \eqref{Ssl2} by expanding
in small momenta $p_k$, $p_j$ of order $\cO(1/L)$ one finds,
\bea
-i \left(
\log S_{\sll(2)}\upper{gauge}(p_k,p_j)-\log S_{\sll(2)}\upper{string}(p_k,p_j)
\right)\simeq\frac{g^4}{2}\,p_k^2\,(p_k-p_j)\,p_j^2+\cO(g^6)\, .
\eea
This proves, incidentally for both the near-BMN limit
as well as the Frolov-Tseytlin limit, two-loop
agreement and three-loop disagreement in the $\sll(2)$ sector,
in full analogy with the $\su(2)$ and $\su(1|1)$ sectors.

\section{Summary and Musing}

Let us collect once more the long-range, asymptotic Bethe ans\"atze we 
proposed for the basic two-component sectors $\su(2)$, $\su(1|1)$ and $\sll(2)$
of $\cN=4$ gauge theory. 
They are all written in the factorized scattering form 
\bea
\label{bethe2}
e^{i p_k L}=
\prod_{\textstyle\atopfrac{j=1}{j\neq k}}^M\, \pm S(p_k,p_j)\, ,
\qquad \qquad
k=1,\ldots,M\, ,
\eea
where the upper sign is for bosons and the lower for fermions.
The three S-matrices are
\bea
\label{proposal1}
S_{\su(2)}\upper{gauge}(p_k,p_j)&=&
\frac{\varphi(p_k)-\varphi(p_j)+i}{\varphi(p_k)-\varphi(p_j)-i}\, , 
\\
S_{\su(1|1)}\upper{gauge}(p_k,p_j)&=& -e^{i \theta(p_k,p_j)}\, ,
\\ \label{proposal13}
S_{\sll(2)}\upper{gauge}(p_k,p_j)&=&
\frac{\varphi(p_k)-\varphi(p_j)-i}{\varphi(p_k)-\varphi(p_j)+i}\,
e^{2 i \theta(p_k,p_j)}\, .
\eea
The phase function $\varphi(p_k)$ is currently known to five loops,
and its all-loop conjecture is \eqref{pf} \cite{BDS}. The 
phase $\theta(p_k,p_j)$ was worked out in section \ref{su11section}
to three-loop order, {\it cf} \eqref{theta},\eqref{thetaalt}. 
It would be fascinating to find its all-loop form.

On the string side, it was found, for $\su(2)$  in \cite{AFS},  and in
section \ref{stringsection} for the other two sectors,
that the near-BMN and Frolov-Tseytlin physics 
is, for large $L$ (i.e.~momenta of order $\cO(1/L)$)
reproduced to all orders in $g^2$ by the S-matrices
\bea
\label{proposal21}
S_{\su(2)}\upper{string}(p_k,p_j)&\simeq&
\frac{\varphi(p_k)-\varphi(p_j)+i}{\varphi(p_k)-\varphi(p_j)-i}\,
\prod_{r=2}^\infty \,
e^{2 i \theta_r(p_k,p_j)}\, , 
\\ \label{proposal22}
S_{\su(1|1)}\upper{string}(p_k,p_j)&\simeq& -e^{-i \theta_1(p_k,p_j)}\, ,
\\ \label{proposal23}
S_{\sll(2)}\upper{string}(p_k,p_j)&\simeq&
\frac{\varphi(p_k)-\varphi(p_j)-i}{\varphi(p_k)-\varphi(p_j)+i}\,
\prod_{r=1}^{\infty}\,
e^{-2 i \theta_r(p_k,p_j)}\, . 
\eea
Finally, it was demonstrated in \cite{B} that to at least five-loop
order a spin chain exists whose S-matrix $\tilde{S}_{\su(2)}$ is
{\it exactly} given by the right hand side of \eqref{proposal21}. 
Interestingly this model is part of the general three-loop 
$\su(2|3)$ spin chain of \cite{dynamic}, which also contains the
$\su(1|1)$ subsector. Therefore supersymmetry predicts that
also a fermionic spin chain should exist which reproduces the string 
and not the gauge theory scattering.
However, as opposed to \eqref{proposal21},
the S-matrix \eqref{proposal22} can unfortunately not, as it stands,
correspond to the one of a quantum spin chain. The reason is
that the shift function $\theta_r(p_k,p_j)$ is not 
periodic \footnote{
It should be clear from the perturbative asymptotic Bethe
ansatz introduced in this paper that the latter will always yield a periodic
S-matrix. The same caveat therefore applies to the $\sll(2)$ string
S-matrix \eqref{proposal23}.}
in the momenta $p_k,p_j$ for $r=1$.  
Excitingly, one may nevertheless find evidence
that the {\it same} deformation which turns the gauge theory spin chain
into the string theory spin chain in the $\su(2)$ sector yields
for the $\su(1|1)$ sector an S-matrix $\tilde{S}_{\su(1|1)}$
which is, to three loop order, ``dressed'' by the same multiplicative
factor $\exp 2 i \theta_2(p_k,p_j)$ as in the $\su(2)$ case.
In fact, we have checked, by including the corresponding deformation 
parameter (denoted by $c_4$ in \cite{B}, where $c_4$ is the gauge
case and $c_4=1$ the string case) into the calculations of 
section \ref{su11section} that the deformation leads to the following 
additional phase shift for $\su(1|1)$
\bea
\label{shift}
\delta \theta(p_1,p_2)= -8\,  c_4 \, 
g^4 \sin^2\left(\frac{p_1}{2}\right)
\Big(\sin p_1-\sin p_2\Big)
\sin^2\left(\frac{p_2}{2}\right)\, ,
\eea
to be added to \eqref{thetaalt}. (Note that in the string case
$c_4=1$ it exactly cancels the last term in \eqref{thetaalt}.) 
This is possible since the
deformation parameter appears in the full three-loop $\su(2|3)$ vertex
of \cite{dynamic}. It is therefore reasonable to
speculate that a spin chain might exists which reproduces the
string results for {\it all} sectors and, maybe, all of
$\psu(2,2|4)$. For the two-component sectors discussed in this
paper this spin chain might have an S-matrix which is related
to the weak-coupling, asymptotic gauge theory S-matrix by the 
same dressing factor:
\bea\label{proposal3}
\tilde{S}_{\su(2)}\upper{string}(p_k,p_j)&=&
S_{\su(2)}\upper{gauge}(p_k,p_j)\,
\hat{S}(p_k,p_j)\, ,
\\ 
\tilde{S}_{\su(1|1)}\upper{string}(p_k,p_j)&=&
S_{\su(1|1)}\upper{gauge}(p_k,p_j)\,
\hat{S}(p_k,p_j)\, , 
\nonumber 
\\
\tilde{S}_{\sll(2)}\upper{string}(p_k,p_j)&=&
S_{\sll(2)}\upper{gauge}(p_k,p_j)\,
\hat{S}(p_k,p_j)\, . 
\nonumber
\eea
where the dressing factor would be approximately given by the product
\bea
\hat{S}(p_k,p_j)\simeq
\prod_{r=2}^\infty \,
e^{2 i \theta_r(p_k,p_j)}\, . 
\eea
Hopefully such a dressing factor will appear in the gauge theory
when we go from weak to strong coupling, e.g.~through the
``wrapping effects'' discussed in \cite{BDS}. 
Or do we have to replace the AdS/CFT ``correspondence'' 
by an AdS/CFT ``similarity''?

It would be
exceedingly important to test these ideas in refined situations such as
$1/J^2$ corrections to the near-BMN limit, and 
$1/L$ corrections to spinning strings. For first steps in
these directions see, respectively,\
\cite{Swanson} and \cite{LZ,FPT}. A dressing factor as in 
\eqref{proposal3}
should then also explain the famous 
$\lambda^{\frac{1}{4}}$ strong coupling behavior. First evidence
that this might happen was found in \cite{AFS}. It would also
be very interesting to reproduce other semiclassical limits,
such as the large spin limit of \cite{GKP2}, from the S-matrices.


\section*{Acknowledgments}

I thank the Kavli Institute for Theoretical Physics for hospitality,
and the organizers of the program
{\it QCD and String Theory} for an inspiring workshop.
I also thank 
Ofer Aharony, Gleb Arutyunov, David Berenstein, Zvi Bern, Vladimir Braun, 
Virginia Dippel, Burkhard Eden, Alexander Gorsky, David Gross, Romuald Janik, 
Martin Kruczenski, Sven-Olaf Moch, Jan Plefka, Joe Polchinski, 
Marcus Spradlin, Matt Strassler, Stefan Theisen, Arkady Tseytlin, 
Anastasia Volovich, Arkady Vainshtein, Kostya Zarembo and, especially, 
Niklas Beisert
for helpful comments, heated discussions, interest,
and useful correspondence.
Thanks to Gleb Arutyunov, Joe Minahan and Kostya Zarembo
for suggestions on the manuscript.
This research was supported in part by the National Science Foundation 
under Grant No. PHY99-07949.






%
\TABLE{
\centering
$
\begin{array}[t]{|l|c c|c|c|l|}\hline
\Delta_0&L&M& \alSU(2|3) &(n_1,\ldots,n_M) &(E_0,E_2,E_4)^{P}\\\hline
4  &3&2&[0;1;0,0]&(-1,1)&(6,-12,42)^+ \\\hline
5  &4&2&[0;1;0,1]&(-1,1)&(4, -6,17)^- \\\hline
6  &5&2&[0;1;0,2]&(-1,1)&(2.76393,-2.90983,6.1067)^+ \\
   & & &         &(-2,2)&(7.23607,-14.0902,52.3933)^+ \\\hline
7  &6&2&[0;1;0,3]&(-1,1)&(2,-\frac{3}{2},\frac{37}{16})^- \\ 
   & & &         &(-2,2)&(6,-\frac{21}{2},\frac{555}{16})^- \\\hline
7  &5&4&[2;3;0,0]&(-2,-1,1,2)&(10,-20,\frac{145}{2})^- \\\hline
7.5&6&3&[1;2;0,2]&\pm(-3,1,2)&(8,-14,49)^\pm \\\hline
8  &7&2&[0;1;0,4]&(-1,1)&(1.50604,-0.830063,0.95726)^+ \\
   & & &         &(-2,2)&(4.89008,-7.30622,20.9555)^+ \\
   & & &         &(-3,3)&(7.60388, -14.8637,57.0872)^+ \\\hline
8  &6&4&[2;3;0,1]&(-2,-1,1,2)&(8,-14,46)^+ \\\hline
8.5&7&3&[1;2;0,3]&\pm(-3,1,2)&(7,-12,\frac{83}{2})^\pm \\\hline
9  &8&2&[0;1;0,5]&(-1,1)&(1.17157,-0.489592,0.432593)^-\\
   & & &         &(-2,2)&(4,-5,\frac{49}{4})^- \\
   & & &         &(-3,3)&(6.82843,-12.5104,44.3174)^- \\\hline
9  &7&4&[2;3;0,2]&(-2,-1,1,2)&(6.39612,-9.3993,27.0234)^- \\
   & & &         &(-3,-1,1,3)&(9.10992,-17.1028,63.4254)^- \\
   & & &         &(-3,-2,2,3)&(12.494,-24.4979,88.5512)^- \\\hline
9.5&8&3&[1;2;0,4]&\pm(-3,1,2)&(6,-\frac{19}{2},\frac{247}{8})^\pm \\
   & & &         &\pm(-4,1,3)&(8,-\frac{29}{2},\frac{427}{8})^\pm \\\hline
\end{array}$
\caption{
The three-loop spectrum of the first few excited states of the 
fermionic subsector $\su(1|1)$ as computed from the explicit
solution \eqref{final}. The latter is derived from
the perturbative Bethe ansatz. The mode numbers entering the
fundamental equation are indicated. The eigenvalues are either
rational or algebraic; in the latter case we have given them
to five digit precision.
In order to compare with the results of direct matrix
diagonalization in \cite{dynamic} (see also p.143 of \cite{thesis})
we have indicated the labels of the corresponding highest weight
state of $\su(2|3)$ (no $\su(1|1)$ state is primary in $\su(2|3)$). 
The agreement is perfect, and establishes perturbative factorized 
scattering in the fermionic sector. Note that the agreement persists
up to large magnon densities, e.g.~$M=4$ magnons in a length $L=5$ spin
chain. $P$ denotes the parity of a state.
}
\label{fermions}
}
%


%
\TABLE{
\centering
$\begin{array}[t]{|c|cc|c|l|}\hline
\Delta_0&L&M&(n_1,\ldots,n_M)& (E_0,E_2,E_4)^{P}\\\hline\hline
4&2&2&(-1,1)&\mathbf{(6,-12,42)^+} \\\hline\hline
5&3&2&(-1,1)&(4,-6,17)^- \\\hline\hline
6&4&2&(-1,1)&(2.76393,-2.90983,6.1067)^+ \\
 & & &(-2,2)&(7.23607,-14.0902,52.3933)^+ \\\hline
6 &3&3&\pm(-2,1,1)&(\frac{15}{2},-\frac{225}{16},\frac{3195}{64})^\pm \\\hline
6 &2&4&(-1,-1,1,1)&
\mathbf{(\frac{25}{3},-\frac{925}{54},\frac{241325}{3888})^+} \\\hline\hline
7&5&2&(-1,1)&(2,-\frac{3}{2},\frac{37}{16})^- \\ 
 & & &(-2,2)&(6,-\frac{21}{2},\frac{555}{16})^- \\\hline
7&4&3&\pm(-2,1,1)&(6,-\frac{21}{2},36)^\pm \\\hline
7&3&4&(-1,-1,1,1)&(6,-\frac{39}{4},\frac{957}{32})^- \\\hline\hline
8&6&2&(-1,1)&(1.50604,-0.830063,0.95726)^+ \\
 & & &(-2,2)&(4.89008,-7.30622,20.9555)^+ \\
 & & &(-3,3)&(7.60388, -14.8637,57.0872)^+ \\\hline
8&5&3&\pm(-2,1,1)&(4.72931,-7.01464,21.1993)^\pm  \\
 & & &\pm(-3,1,2)&(7.77069,-14.4229,52.8944)^\pm  \\\hline
8&4&4&(-1,-1,1,1)&(4.38277,-5.25026,12.58394)^+  \\
 & & &(-2,-1,1,2)&(8.35923,-16.0680,59.3810)^+  \\
 & & &(-2,-2,2,2)&(11.5913,-23.1031,83.6199)^+  \\
 & & &\pm(1,1,1,1)&(\frac{23}{3},-\frac{1331}{108},
\frac{76973}{1944})^\pm \\\hline
8&3&5&\pm(-2,-1,1,1,1)&(\frac{35}{4},-\frac{18865}{1152},
\frac{1068515}{18432})^\pm \\\hline
8&2&6&(-1,-1,-1,1,1,1) 
&\mathbf{(\frac{49}{5},-\frac{45619}{2250},\frac{300642097}{4050000})^+} \\\hline
\end{array}$
\caption{
The three-loop spectrum of the first few highest weight states of the 
derivative subsector $\sll(2)$ as derived from the
perturbative Bethe ansatz.  The mode numbers entering the
fundamental equation are indicated. The eigenvalues are either
rational or algebraic; in the latter case we have given them
to five digit precision. The one-loop spectrum was found by
direct matrix diagonalization in
\cite{complete} (see also p.84 of \cite{thesis}).
At the time of writing this spectrum
cannot be checked against direct diagonalization since the
$\sll(2)$ dilatation operator is not yet known beyond one loop.
The bold-faced values with $L=2$ correspond to twist-two operators and
agree perfectly with results of 
Kotikov, Lipatov, Onishchenko and Velizhanin. The values of two-magnon 
states $M=2$ were previously known from superconformal invariance.
The three-loop results for the cases $L=2,3$ and $M=2$ were
first predicted in \cite{BKS} and confirmed in a field theory
computation in \cite{EJS}.
The two- and three loop anomalous dimensions of all twist-$L$ operators 
with $L>2$ and $M>2$ are new predictions.
The parity of a state is denoted by $P$.
The case $M=4$, $L=4$ with mode numbers $\pm(1,1,1,1)$
is interesting as it is the simplest example of a paired state in
$\sll(2)$ where all Bethe roots are either strictly positive or
strictly negative real. Such states with ``winding number'' were 
considered, in the thermodynamic limit, in \cite{smedback,KZ}, and
it is reassuring to find them in the finite spin chain. 
}
\label{derivatives}
}

\end{document}